\documentclass[8pt,newlfont,aps,pra,notitlepage,twocolumn,superscriptaddress,showpacs]{revtex4}
\def\hs{\hspace{0.5cm}}
\def\ft{\hspace{1mm}}
\def\ea{{\it et al.}}
\usepackage{color}
\usepackage{graphicx}
\usepackage{exscale}
\usepackage{amsfonts}
\usepackage{sidecap}
\begin{document}
\title{Properties of bosons in a one-dimensional bichromatic optical lattice
in the regime of the Sine-Gordon transition: a Worm Algorithm Monte Carlo study}
\author{Asaad R. Sakhel}
\affiliation{Department of Physics and Basic Sciences, Faculty of Engineering Technology, Al-Balqa Applied 
University, Amman 11134, Jordan}
\affiliation{Abdus-Salam International Center for Theoretical Physics, Strada Costiera 11, 34151 Trieste, Italy}
\begin{abstract}
The sensitivity of the Sine-Gordon (SG) transition of strongly interacting bosons confined in a shallow, 
one-dimensional (1D), periodic optical lattice (OL), is examined against perturbations of the OL. The SG 
transition has been recently realized experimentally by Haller \ea\ [Nature $\mathbf{466}$, 597 (2010)] and 
is the exact opposite of the superfluid (SF) to Mott insulator (MI) transition in a deep OL with weakly 
interacting bosons. The continuous-space worm algorithm (WA) Monte Carlo method [Boninsegni \ea, Phys. Rev. 
E $\mathbf{74}$, 036701 (2006)] is applied for the present examination. It is found that the WA is able to 
reproduce the SG transition which is another manifestation of the power of continuous-space WA methods in 
capturing the physics of phase transitions. In order to examine the sensitivity of the SG transition, it is 
tweaked by the addition of the secondary OL. The resulting bichromatic optical lattice (BCOL) is considered 
with a rational ratio of the constituting wavelengths $\lambda_1$ and $\lambda_2$ in contrast to the commonly 
used irrational ratio. For a weak BCOL, it is chiefly demonstrated that this transition is robust against the 
introduction of a weaker, secondary OL. The system is explored numerically by scanning its properties in  
a range of the Lieb-Liniger interaction parameter $\gamma$ in the regime of the SG transition. It is argued 
that there should not be much difference in the results between those due to an irrational ratio $\lambda_1/\lambda_2$ 
and due to a rational approximation of the latter, bringing this in line with a recent statement by 
Bo$\acute{e}$ris \ea\ [Phys. Rev. A {\bf 93}, 011601(R) (2016)]. The correlation function, Matsubara Green's 
function (MGF), and the single-particle density matrix do not respond to changes in the 
depth of the secondary OL $V_1$. For a stronger BCOL, however, a response is observed because of changes 
in $V_1$. In the regime where the bosons are fermionized, the MGF reveals that hole excitations are favored 
over particle excitatons manifesting that holes in the SG regime play an important role in the response of 
properties to changes in $\gamma$.  
\end{abstract}
\date{\today}
\maketitle

\section{Introduction}\label{sec:introduction}

\hs It is known that weakly interacting bosons confined by a deep periodic one-dimensional (1D) optical lattice (OL) undergo 
a transition from a superfluid to a Mott insulator (MI) \cite{Sengupta:2005,Liu:2005,Clark:2004,Ramanan:2009} when the 
strength of the OL is increased. The opposite situation of getting another kind of SF to MI transition in a shallow 
1D OL occupied by strongly interacting bosons \cite{Haller:2010,Buechler:2003} is known as the Sine-Gordon (SG) transition, 
where the bosons are ``pinned" just by subjecting them to a very weak periodic OL. Indeed, the SG transition has been just 
recently realized experimentally \cite{Haller:2010}, although its concept has been communicated and examined earlier 
\cite{Giamarchi:2003,Buechler:2003} even in a hollow-core 1D fibre \cite{Huo:2012}. In the phase diagram Fig.4 of Haller 
\ea\ \cite{Haller:2010}, the SG regime exists for interactions $\gamma\ge\gamma_c=3.5$ and a vanishingly shallow 1D OL of 
depth $V_0<E_R$ where $E_R$ is the recoil energy and $\gamma$ ($\gamma_c$) the Lieb-Liniger (critical) interaction parameter 
\cite{Lieb:1963}.

\hs The main goal of this article is the examination of the sensitivity of this pinning transition against perturbations. 
In that sense, we tweak the SG transition by the addition of a second disordering OL component; a method that has been 
used elsewhere \cite{Guidoni:1997,Nessi:2011,Gordillo:2015}. Since the latter transition has been observed in a periodic 
OL \cite{Haller:2010}, it is justified to explore its stability under a perturbation of the latter OL. In fact the chief 
result of this work is that the SG transition is robust against the latter perturbations. This result may turn out to
be important in the research on hollow-core 1D fibres filled with ultracold atoms, where photons have been pinned
by a shallow effective polaritonic potential just like atoms \cite{Huo:2012} and the quantum transport of strongly
interacting photons \cite{Hafezi:2012}.

\hs Another goal is the examination of the properties of the bosons as measured by the pair correlation 
function $g_2(r)$ and one-body density matrix (OBDM) $g_1(r)$ (with $r$ the distance between a pair of particles), the Matsubara 
Green's function (MGF) $G(p=0;\tau)$, and superfluid fraction $\rho_s/\rho$ in the regime of this transition. Both of 
$g_2(r)$ and $g_1(r)$ are spatially averaged quantities as given below by Eqs.(\ref{eq:g2r}) and (\ref{eq:obdm-homogeneous}), 
respectively. It must be emphasized that in this work the $g_2(r)$ of the homogeneous Bose gas has been intentionally applied 
to the bosons in a 1D bichromatic optical lattice (BCOL), for mathematical convenience, to account for the density-density 
correlations under the effect of a BCOL. The pair correlations are inherently related to, and give an account of, 
the density-density correlations \cite{Pethick:2002,Lewenstein:2012}. On the other hand for inhomogeneous systems, the pair 
correlation function and OBDM are normalized by convolutions of the spatially varying density instead of the average 
linear density (a constant). These are given by what is equal to $h_2(r)=g_2(r)n_0^2/\rho_c(r)$ and 
$h_1(r)=g_1(r)n_0/\rho_{c,\frac{1}{2}}(r)$ in Eqs.(\ref{eq:normcorrh2r}) and (\ref{eq:obdm-inhomogeneous}), respectively.
Here $\rho_c(r)$ and $\rho_{c,\frac{1}{2}}(r)$ are defined by Eqs.(\ref{eq:spatially-varying-density-norm}) and (\ref{eq:rho-half}), 
respectively, and $n_0$ is the average linear density. It is found that one can obain from $g_2(r)$ appreciable signals arising 
from a strong response to changes in the primary OL and interactions. Further, $g_1(r)$ applied to the BCOL system displays 
almost the same behavior as $h_1(r)$, except for some small differences. In $h_2(r)$, the normalization by $\rho_c(r)$ is 
found to weaken these signals substantially. A key result is that for a shallow 1D OL, the above properties are 
not influenced by perturbations of the OL, bringing this in line with a recent examination \cite{Boeris:2016}. We particularly 
focus on $g_2(r)$ because it is important to examine it with regards to this and other kinds of phase transitions. For example 
the pair correlation function for a 1D uniform Bose gas has been used as the ratio between the photoassociation rates of 
Rb$^{87}$ atoms in 1D and 3D \cite{Kinoshita:2005}. Although we examine it only briefly, the $g_1(r)$ is also not any less 
important; for example Deissler \ea\ \cite{Deissler:2011} presented the first experimental analysis of a spatially averaged 
$g_1(r)$, similar to ours, in a quasiperiodic optical lattice (QPOL).

\hs Remaining within the regime of the SG--transition, we explore it in a BCOL instead 
of a 1D OL to study its sensitivity --in addition to that of the associated properties-- to changes in the BCOL. Indeed 
the role of a 1D OL, such as the BCOL \cite{Guidoni:1997,Nessi:2011}, and in conjunction with atom-atom interactions in 
defining the properties of confined bosons lies at the heart of many investigations today 
\cite{Gordillo:2015,Kinoshita:2005,Roux:2008,Roscilde:2008,Larcher:2011,Modugno:2009,Roth:2003}. 
So far, the BCOL has been mostly applied to introduce quasidisorder in a ``common experimental route" \cite{Gordillo:2015}. 
This is usually achieved by superimposing two OL wavelengths whose ratio $\lambda_1/\lambda_2$ yields an irrational number 
\cite{Modugno:2009,Roux:2008,Larcher:2011}. However, the lattice setup with a rational number $\lambda_1/\lambda_2$ is not 
very common and deserves therefore an investigation, particularly due to the likelihood that there may be not much 
difference between the use of a rational and irrational $\lambda_1/\lambda_2$. The same argument has been made recently by 
Boeris \ea\ \cite{Boeris:2016} who stated (quoting them): ``... it is not necessary to implement truly irrational numbers
with mathematical (i.e., unattainable) precision; after all, on a finite lattice one can ``resolve" only a finite
number of digits." In fact, real disorder can only be achieved by a speckle potential and the investigation of bosons in 
this kind of potential, and add to this a quasidisordered one, has been going on intensively in the last few years 
\cite{Nessi:2011,Pilati:2010,Aleiner:2010,Deissler:2011,Lugan:2009,White:2009,Fallani:2007,Roscilde:2008,
Deissler:2010,Roati:2008,Billy:2008,Chen:2008,Fisher:1989,Bossy:2008,Palpacelli:2008,Clement:2006,Deng:2013,
Pollet:2013,Ristivojevic:2012,Iyer:2013,Basko:2013,Schulte:2005,Paul:2007,Sanchez-Palencia:2007,Lugan:2007,
Paul:2009,Radic:2010,Cestari:2010,Iyer:2012,Aulbach:2004,Boers:2007,Modugno:2009,Deng:2014}. The minor side 
issue on the difference between the results due to rational and irrational ratios $\lambda_1/\lambda_2$ will be 
discussed briefly in a section later on here, where results are given showing that it makes no difference.

\hs An investigation that is very relevant to the present work is that by Gordillo \ea\ \cite{Gordillo:2015} 
who calculated the phase diagram of a continuous system of bosons in a BCOL. By keeping the interaction strength 
fixed, the superfluid fraction has been examined as a function of the secondary-OL depth for several values of the 
primary-OL depth within a range of amplitudes which is larger than ours. It must be emphasized that these authors 
used a rational approximation to an {\it irrational} ratio of $\lambda_1$ and $\lambda_2$ and it has been argued 
that this is ``common practice" \cite{Roth:2003,Roscilde:2008}. Among their findings, it has been demonstrated 
that changes in the secondary OL influence the properties and that an MI can be realized. In contrast, the present 
work conducts a similar investigation chiefly by varying the interaction strength in a {\it shallow} BCOL with a 
{\it rational} $\lambda_1/\lambda_2$ in a regime of interactions near the SG transition, although it is argued 
that this ratio can also arise from an approximation to the irrational one. It has been found that under these 
conditions, the properties of bosons are not influenced by changes in the intensity of the secondary OL. Our studies 
therefore complement the work of Gordillo \ea, and are further substantiated with the examination of 
the SG transition. Moreover, for a strong BCOL additionally considered, changes in the properties can be observed 
as one varies the intensity of the secondary OL bringing this is in line with the results of Gordillo \ea\ 
\cite{Gordillo:2015}.

\hs The bosons in the BCOL are simulated using the continuous$-$space worm algorithm (WA) quantum Monte Carlo 
approach \cite{Boninsegni:2006}. It is found that WA reproduces the SG transition accurately and that the interplay 
of BCOL and interactions has little effect on changing the critical interaction at which this transition occurs. In 
fact, it shall be reasoned below that the same result could have been obtained by using irrational ratios close to the 
rational ones. In other results (1) the OBDM of the system displays substantial depletion of the superfluid as it 
passes through the SG transition; (2) the MGF shows signals for fermionization detected via the correlation 
function at the origin, $g_2(0)$, when the total interaction energy goes to zero demonstrating perfect antibunching 
\cite{Sykes:2008}; (3) the secondary OL has been found not to play a role in aiding or preventing the fermionization. 

\hs This work provides more clarification for the interplay between interactions and disorder for repulsive bosons 
in a quasidisordered OL as has been given earlier by Deissler \ea\ \cite{Deissler:2010}. It is believed that this 
research will be an important contribution to the field of disordered bosons in 1D. 

\hs The organization of the present paper is as follows. In Sec.~\ref{sec:method} the method is presented.
In Sec.~\ref{sec:tests-of-WA-code} the WA code is tested for accuracy as applied to the present system.
In Sec.~\ref{sec:results} the results are presented. In Sec.~\ref{sec:rational-vs-irrational} the issue of 
a rational and irrational ratio of the BCOL wavelengths is discussed. In Sec.\ref{sec:conclusion} the paper 
concludes with some closing remarks. In Appendix \ref{app:wpimc} the WA is briefly described. In Appendices
\ref{app:norm-corr-function-density}, \ref{app:sa-obdm}, and \ref{app:local-h2(0)}, the $h_2(r)$, $h_1(r)$,
and $h_2(r=0)$, given by Eqs.(\ref{eq:normcorrh2r}) and (\ref{eq:obdm-inhomogeneous}) below, are reexamined 
for the same systems in Figs.\ft\ref{fig:plotdensmcorrA1.579severalBasc2.000T0.001L500N200}, 
\ref{fig:plotcorrGCANseveralAandBT0.001den0.400L500N200severalGammaStack},
and \ref{fig:compareAnalyticalcorr0withWAPIMC} that display $g_2(r)$, $g_1(r)$ and $g_2(0)$, respectively,
given by Eqs.(\ref{eq:g2r}) and (\ref{eq:obdm-homogeneous}).

\section{Method}\label{sec:method}

\hs A brief description of the WA applied to the present system is relegated to Appendix \ref{app:wpimc}. 
The simulations have been conducted on the excellent computational cluster of the Max Planck Institute for 
Physics of Complex Systems in Dresden, Germany. In essence, the present work has been a heavy-computational 
project with each simulation taking about a week of CPU time to finish.

\subsection{Optical lattice}

\hs In order to introduce disorder into the present system, the WA code \cite{Prokofev:2011} --originally 
designed for a homogeneous Bose gas-- has been modified by including a BCOL potential of the form

\begin{equation}
V_{OL}(x)\,=\,V_0\cos^2(\alpha\pi x)\,+\,V_1\cos^2(\beta\pi x),
\label{eq:quasiperiodic-optical-lattice}
\end{equation}

where $V_0$ and $V_1$ are the primary and secondary OL depths, respectively, and we always consider $V_1<V_0$. 
The parameters $\alpha=2/\lambda_1$ and $\beta=2/\lambda_2$, with $\lambda_1$ and $\lambda_2$ the wavelengths, 
determine the periodicity of the BCOL and such a type of lattice was shown to realize disorder 
\cite{Fallani:2007,Nessi:2011}. $\alpha$ and $\beta$ were set to 0.4 and 1.0, respectively, for a QPOL, and 
0.4 and 1.39 for a quasidisordered optical lattice (QDOL). 

\hs A measure for the the strength of the BCOL quasidisorder is its standard deviation $\delta V$ given by 

\begin{equation}
\delta V\,=\,\sqrt{\langle V^2\rangle\,-\,\langle V\rangle^2},
\label{eq:deltaV}
\end{equation}

with 

\begin{equation}
\langle V\rangle\,=\,\frac{1}{L}\,\int_0^L V_{OL}(x) dx,
\label{eq:average-V}
\end{equation}

and 

\begin{equation}
\langle V^2\rangle\,=\,\frac{1}{L}\,\int_0^L V_{OL}^2(x) dx.
\label{eq:average-Vsq}
\end{equation}

According to Deissler \ea\ \cite{Deissler:2011}, another measure for the disorder strength
is given by 

\begin{equation}
\Delta\,=\,0.5 V_1 \epsilon^2 \exp(-2.18/V_0^{0.6}),
\label{eq:delta-disorder}
\end{equation}

and considering the tunneling amplitude between adjacent lattice sites 

\begin{equation}
J\,=\,1.43 V_0^{0.98}\exp(-2.07\sqrt{V_0}),
\label{eq:J-tunneling-amplitude}
\end{equation}

this allows one to evaluate $\Delta/J$. In (\ref{eq:delta-disorder}) and (\ref{eq:J-tunneling-amplitude}), 
$V_0$ and $V_1$ are in units of $E_R$ (see Sec.\ref{sec:units-and-parameters}), and $\epsilon=\beta/\alpha$ 
is the ratio between the wavelengths of the BCOL.

\subsection{Interactions}
 
\hs The interactions between the bosons are accounted for by the exact two-particle density matrix 
given by (see e.g. Feynman \cite{Feynman:1998})

\begin{eqnarray}
&&\rho(x_1-x_2)\,=\,\nonumber\\
&&1\,-\,\sqrt{\frac{\tau}{4 m a_s^2}}\,
\exp\left\{\frac{m}{4 \tau}(x_1-x_2)^2\,+\,\frac{|x_1|+|x_2|}{a_{1D}}\,+\,\right.\nonumber\\
&&\left.\frac{\tau}{m a_{1D}^2} \right\}, 
\label{eq:two-particle-density-matrix}
\end{eqnarray}

where $\tau=\beta/M$ is the ``time step", with $\beta=1/(\widetilde{T} T_d)$, $\widetilde{T}$ being 
the temperature in units of the degeneracy temperature $T_d$, $k_B=1$ Boltzmann's constant, and $m=0.5$
is the boson mass. Eq.(\ref{eq:two-particle-density-matrix}) appears in the worm-update probabilities as 
a multiplicative factor. The interactions are then essentially described by a delta function 
$g_{1D}\delta(x_1-x_2)$, where $g_{1D}$ is the interaction parameter. 

\hs In Astrakharchik \ea\ \cite{Astrakharchik:2005}, $g_{1D}$ is given by

\begin{equation}
g_{1D}\,=\,-\frac{2\hbar^2}{m a_{1D}},
\label{eq:relation.a1D.and.a3D}
\end{equation}

where $\hbar$ is Planck's constant and $a_{1D}$ the scattering length. Note that although $g_{1D}$ is 
negative, it does not indicate attraction which in itself is counterintuitive. Essentially, the absolute 
value of $g_{1D}$ is considered in the present calculations. From Haller \ea\ \cite{Haller:2010}, the 
Lieb-Liniger parameter is given by $\gamma = m g_{1D}/(\hbar^2 n_0)$, and therefore from 
Eq.(\ref{eq:relation.a1D.and.a3D}) one gets $\gamma\,=\,-2/(a_{1D} n_0)$. Here $n_0=\langle N\rangle/L$ is 
the average linear density of the system with $\langle N\rangle$ the thermodynamic average of the particle 
number and $L$ the length of the system.

\subsection{Pair correlation function}\label{sec:pair-correlation}%Method Section

\hs The spatially-averaged pair correlation function (SAPCF) for the 1D homogeneous Bose gas 
is given by

\begin{equation}
g_2(r)\,=\,\frac{1}{L\,n_0^2}\int_0^L dx\,\langle\hat{\psi}^\dagger(x+r) \hat{\psi}^\dagger(x) 
\hat{\psi}(x)\hat{\psi}(x+r)\rangle. \label{eq:g2r}
\end{equation}

where $r$ is the distance between any two particles along the system and $\hat{\psi}(x)$ [$\hat{\psi}^\dagger(x)$] 
the field operator annihilating [creating] a boson at position $x$. It can account for the spatially averaged 
density-density correlation function, given by

\begin{equation}
n_c(r)\,=\,\frac{1}{L}\int_0^L dx\,\langle \hat{n}(x+r) \hat{n}(x)\rangle 
\label{eq:density-density-correlation}
\end{equation}

where $\hat{n}(x)=\hat{\psi}^\dagger(x)\hat{\psi}(x)$ is the density operator. This is because the pair
correlations $\langle\hat{\psi}^\dagger(x+r)\hat{\psi}^\dagger(x)\hat{\psi}(x)\hat{\psi}(x+r)\rangle$ are
obtainable from the density-density correlations $\langle\hat{n}(x+r)\hat{n}(x)\rangle$ via the
commutator $[\hat{\psi}(x),\hat{\psi}^\dagger(x')]\,=\,\delta^{(3)}(x-x')$ \cite{PethickEqn}

\begin{eqnarray}
&&\langle\hat{\psi}^\dagger(x+r)\hat{\psi}(x+r)\hat{\psi}^\dagger(x)\hat{\psi}(x)\rangle\,=\,\nonumber\\
&&\langle\hat{\psi}^\dagger(x+r)\hat{\psi}^\dagger(x)\hat{\psi}(x+r)\hat{\psi}(x)\rangle\,+\,
\langle\hat{\psi}^\dagger(x+r)\hat{\psi}(x)\rangle\delta^{(3)}(r)\nonumber\\ \label{eq:relation-den-den-corr-pair-corr}
\end{eqnarray}

so that

\begin{equation}
n_c(r)\,=\,g_2(r)n_0^2+\frac{1}{L}\delta^{(3)}(r)\,\int_0^L dx \langle\hat{\psi}^\dagger(x+r)\hat{\psi}(x)\rangle.
\label{eq:nc(r)-g2(r)}
\end{equation}

It must be noted that for a translationally invariant system, Eq.(\ref{eq:g2r}) reduces effectively
to $\langle\hat{\psi}^\dagger(x+r) \hat{\psi}^\dagger(x) \hat{\psi}(x)\hat{\psi}(x+r)\rangle/n_0^2$ and is
the same for all $x$. Therefore by integrating over all $x$ for a translationally invariant system, we 
average the latter over all the variations with $x$. On the other hand, when $g_2(r)$ is normalized by 
the convolution of two spatially-varying densites, $\rho(x)$ and $\rho(x+r)$, that is

\begin{equation}
\rho_c(r)\,=\,\frac{1}{L}\int_0^L dx \rho(x+r)\rho(x)
\label{eq:spatially-varying-density-norm}
\end{equation}

instead of $n_0^2$, where $\rho(x)=\langle\hat{\psi}^\dagger(x)\hat{\psi}(x)\rangle$, one gets the SAPCF 
for an inhomogeneous BEC given by \cite{footnote1,Naraschewski:1999}

\begin{equation}
h_2(r)\,=\,\frac{\displaystyle\int_0^L dx\,\langle\hat{\psi}^\dagger(x+r) \hat{\psi}^\dagger(x) 
\hat{\psi}(x)\hat{\psi}(x+r)\rangle}{\displaystyle\int_0^L dx\,\rho(x+r)\rho(x)}.
\label{eq:normcorrh2r}
\end{equation}

The examination of $h_2(r)$ is relegated to Appendix \ref{app:norm-corr-function-density}. 

\subsection{One-body density matrix}\label{sec:sa-obdm}

\hs The spatially-averaged one-body density matrix (SA-OBDM) for the 1D homogeneous Bose gas is 
defined as

\begin{equation}
g_1(r)\,=\,\frac{1}{L\,n_0}\displaystyle\int_0^L dx\,\langle\hat{\psi}^\dagger(x+r)\hat{\psi}(x)\rangle,
\label{eq:obdm-homogeneous}
\end{equation}

and again for the inhomogeneous case \cite{Naraschewski:1999}

\begin{equation}
h_1(r)\,=\,\frac{\displaystyle\int_0^L dx\,\langle\hat{\psi}^\dagger(x+r)\hat{\psi}(x)\rangle}
{\rho_{c,\frac{1}{2}}(r)}.
\label{eq:obdm-inhomogeneous}
\end{equation}

where 

\begin{equation}
\rho_{c,\frac{1}{2}}(r)\,=\,\frac{1}{L}\displaystyle\int_0^L dx \sqrt{\rho(x+r)\rho(x)}
\label{eq:rho-half}
\end{equation}

Eq.(\ref{eq:obdm-homogeneous}) is used with the same purpose of comparison as in Sec.(\ref{sec:pair-correlation}).
The examination of $h_1(r)$ is also relegated to Appendix \ref{app:sa-obdm}. $\rho_c(r)$ and $\rho_{c,\frac{1}{2}}(r)$ 
will be from now on referred to as the ``convoluted densities". In passing, it must be noted that the
normalization by $n_0^2$ in $g_2(r)$, and by $\rho_c(r)$ in $h_2(r)$, just remains a matter of convenience
regarding what purpose each one would achieve. For $g_2(r)$ this simply entails information about the
density-density correlations and stronger signals than $h_2(r)$. For $g_1(r)$ and $h_1(r)$ it turns out 
that the normalization by $n_0$ and $\rho_{c,\frac{1}{2}}(r)$, respectively, doesn't make much difference. It must 
be added that experimentally one favors measuring the spatially-integrated correlation functions, Eqs.(\ref{eq:g2r}) 
and (\ref{eq:obdm-homogeneous}), because otherwise it requires a lot of data to determine the dependence of the 
correlations on two coordinates, a fact that has been stated earlier by Naraschewski et al. \cite{Naraschewski:1999}.

\subsection{Units and parameters}\label{sec:units-and-parameters}

\hs The degeneracy temperature is given by $T_d\,=\,2\pi\hbar^2 n_0^2/m$, and the photon recoil energy by
$E_R\,=\,h^2/(2 m \lambda^2)$; their ratio is always $T_d/E_R\,=\,4/\pi$. Here $\lambda$ is the wavelength 
of the laser beams creating the OL where $\lambda=2d$ is twice the lattice period $d$ [$=1/\alpha$; cf. 
Eq.(\ref{eq:quasiperiodic-optical-lattice})] and $n_0=2/\lambda$ is the average linear density. Hence, an OL depth like, 
e.g., 1.0 in units of $E_R$ corresponds to $V_0\,=\,1.0\,E_R/T_d\,=\,1.0 \times \pi/4\,=\,0.785$ in units 
of $T_d$. Now in the WA code the mass is $m=0.5$, $\hbar=1$, $\lambda/L=0.01$, and $n_0\lambda=2.0$, and 
therefore $T_d\,=\,4\pi n_0^2\,=\,0.64 \pi$ and $E_R\,=\,4\pi^2 \hbar^2/(2 m\lambda^2)\,=\,0.16 \pi^2$. It 
should be emphasized that $T_d$ is computed from an initial density parameter $n_0$ that is only used to fix 
the size of the system and that is not updated during the simulations. That is, the final density of the 
system $n$ can be different from the initial $n_0$. However, in the present simulations $n_0$ is equal to 
$n$ and the temperature is set to $\widetilde{T}=T/T_d=0.001$ to allow a significant value of $\rho_s/\rho$. 

\hs The length of the system is such that $2L/\lambda\,=\,200$ lattice sites and is at almost perfect commensurate 
filling. For purposes of comparison, the value of the wavelength is the same as that used by Haller \ea\ 
\cite{Haller:2010}, namely $\lambda=1064.5\,\hbox{nm}$ and the values for the OL depth $V_0$ are of the same order 
of magnitude as in Haller \ea\ \cite{Haller:2010} and Gordillo \ea\ \cite{Gordillo:2015}. In \cite{Gordillo:2015}, 
the wavelength of the laser beams generating the OL was $\lambda=830\,\hbox{nm}$ and considering that the 3D scattering 
length of Rb${}^{87}$ is almost $100 a_0$, the 1D Lieb-Liniger interaction parameter becomes $\gamma=1.770$. The 
values of $\gamma$ used in the present work are in general larger than in Ref.~\cite{Gordillo:2015} as they extend 
into the strongly interacting SG regime.

\begin{figure}[t!]
\includegraphics[width=9.0cm,bb=64 413 310 701,clip]{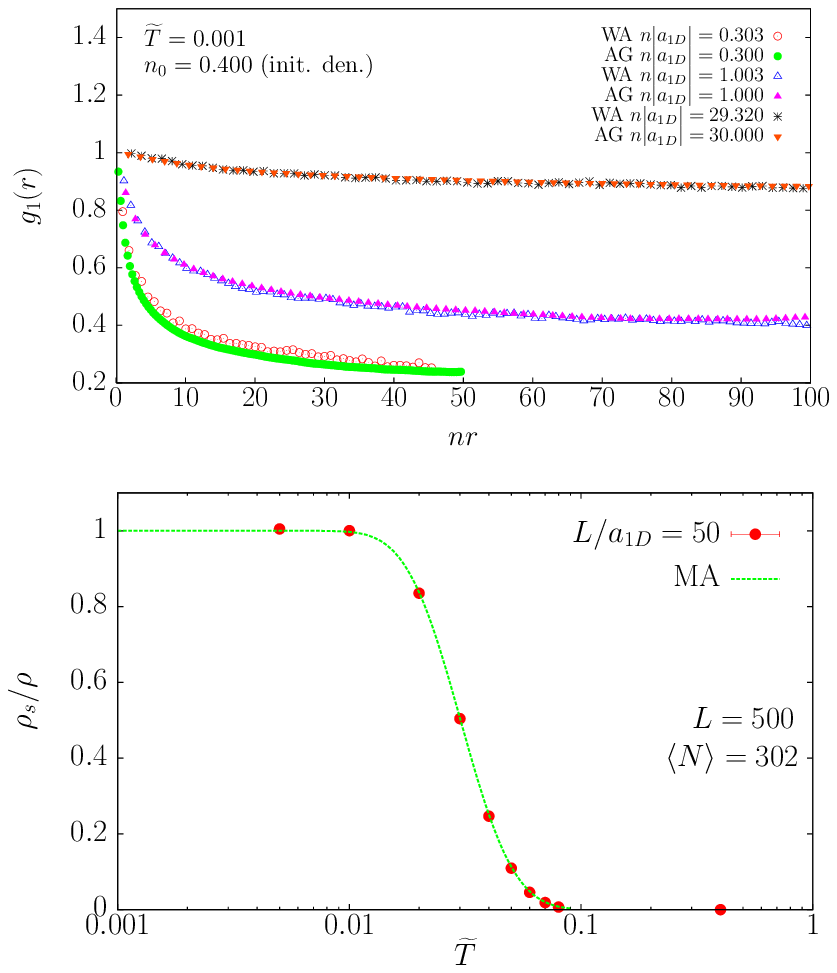}
\caption{(Color online) Upper frame: WA first order correlation function $g_1(r)$ (one-body density matrix) 
as a function of $nr$ for several parameters $n|a_{1D}|$ compared to results from Astrakharchik and Giorgini 
(AG) \cite{Astrakharchik:2005} for values of $n|a_{1D}|$ which are almost identical to the WA ones. The 
system is a 1D homogeneous Bose gas at a temperature of $\widetilde{T}=0.001$. Here $n$ is the average 
linear density and $a_{1D}$ the 1D scattering length. For the WA we have: $n|a_{1D}|=0.303$ (open circles), 
1.003 (open up triangles), and 29.320 (stars), respectively. The corresponding data from AG are for almost 
the same $n|a_{1D}|$: 0.300 (solid circles), 1.000 (solid up triangles), and 30.000 solid down triangles, 
respectively. $\widetilde{T}$ is in units of the transition temperature $T_d$ and the initial density 
of the simulation is $n_0=0.4$. Lower frame: Superfluid fraction $\rho_s/\rho$ as a function of temperature 
$\widetilde{T}$. The system is again a 1D homogeneous gas of Bosons. The scattering length is $a_{1D}/L=1/50$, 
the length of the system is $L/a_{1D}=50$ and the thermodynamic average of the number of particles is 
$\langle N\rangle=302$. The solid circles are the WA results whereas the solid line is an analytical 
calculation using Eq.(\ref{eq:DelMaestroAffleckSFF}) of Del Maestro and Affleck \cite{DelMaestro:2010} 
with the same parameters as ours (see text).}
\label{fig:plot.obdm.vs.nr.and.sff.vs.T.two.figures}
\end{figure}

\section{Tests of the WA code}\label{sec:tests-of-WA-code}

\hs In this section the WA code \cite{Prokofev:2011} is tested on a uniform interacting Bose gas in the 
absence of any trapping potential. First, it is verified that the code produces the OBDM properly via a 
comparison with previous results. Second, it is confirmed that for suitably chosen parameters, the temperature 
of the simulations is low enough to obtain a significant superfluid fraction $\rho_s/\rho$. In this respect, 
WA results for $\rho_s/\rho$ as a function of $\widetilde{T}$ were found to exactly match an analytical 
calculation. In another decisive program check, although this is implied in the subsequent Sec.~\ref{sec:SG-transition}, 
the SG transition originally observed by Haller \ea\ \cite{Haller:2010} was reproduced by adding a shallow 
periodic 1D OL to the homogeneous Bose-gas system.

\subsection{One-body density matrix}

\hs The upper frame of Fig.\ft\ref{fig:plot.obdm.vs.nr.and.sff.vs.T.two.figures} displays 
the OBDM $g_1(r)$ as a function of $nr$ at various $n|a_{1D}|$ obtained from WA simulations for three 
of the homogeneous Bose-gas cases already considered by Astrakharchik and Giorgini (AG) \cite{Astrakharchik:2005}. 
The values of $n|a_{1D}|$ range from the strongly to the weakly-interacting regime as $n|a_{1D}|$
is increased. One can see that the agreement is excellent.

\subsection{Superfluid fraction}

\hs The lower frame of Fig.\ft\ref{fig:plot.obdm.vs.nr.and.sff.vs.T.two.figures} displays a 
comparison between the superfluid fraction $\rho_s/\rho$ obtained by WA and that by the equation

\begin{equation}
\rho_s/\rho\,=\,1\,-\,u\left|\frac{\theta_3^{\prime\prime}(0,e^{-2\pi u})}{\theta_3(0,e^{-2\pi u})}\right|
\label{eq:DelMaestroAffleckSFF}
\end{equation}

that is the same as Eq.(16) of Ref.\cite{DelMaestro:2010}; except that it is rescaled to our units. 
Here $u=1/(2\widetilde{T} \langle N\rangle)$, with $\widetilde{T}=T/T_d$, $\langle N\rangle\,=\,302$, 
and $L/a_{1D}=50$. $\theta_3$ is the Jacobi Theta function of the third kind given by

\begin{equation}
\theta_3(z,q)\,=\,\sum_{n=-\infty}^{+\infty} q^{n^2} \exp(i 2 n z),
\label{eq:theta3-function}
\end{equation}

and can be evaluated using WOLFRAM MATHEMATICA${}^\textsuperscript{\textregistered}$. The system is a 
weakly interacting, dilute, and uniform 1D Bose gas whose initial density was set to $n_0=0.2$ so that 
$T_d=0.5027$. The WA results match exactly those of the analytical Eq.(\ref{eq:DelMaestroAffleckSFF}) 
casting away all doubts about the accuracy of the WA code. Further, $\widetilde{T}=0.001$ is low enough 
to allow a significant value for $\rho_s/\rho$. The reader must be alerted, that the parameters used in
this section and the previous one are only for the purpose of making the comparisons in 
Fig.\ft\ref{fig:plot.obdm.vs.nr.and.sff.vs.T.two.figures}. The rest of this paper uses the parameters of 
Sec.\ft\ref{sec:units-and-parameters} above.

\begin{figure}[t!]
\includegraphics[width=8.5cm,bb=57 397 525 682,clip]{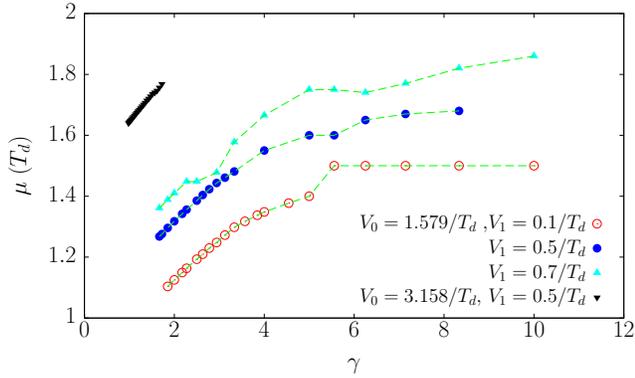}
\caption{(Color online) Calibrated chemical potential $\mu$ as a function of $\gamma$ for the commensurate filling
of a number of BCOLs with $\alpha=0.4$ and $\beta=1.0$ at $N=200\pm\delta$ particles, where $\delta$ 
is a small error. The figure displays the case for $V_0=1.579/T_d$ and $V_1=0.1/T_d$ (open circles); 
$0.5/T_d$ (solid circles); $0.7/T_d$ (solid up triangles); and then the case for $V_0=3.158/T_d$ and 
$V_1=0.5/T_d$ (solid down triangles). $\mu$ is in units of $T_d$ and $\gamma$ is unitless.}
\label{fig:plot.mu.vs.gamma.several.A.and.B}
\end{figure}

\begin{figure}[t!]
\includegraphics[width=8.8cm,bb=61 451 331 742,clip]{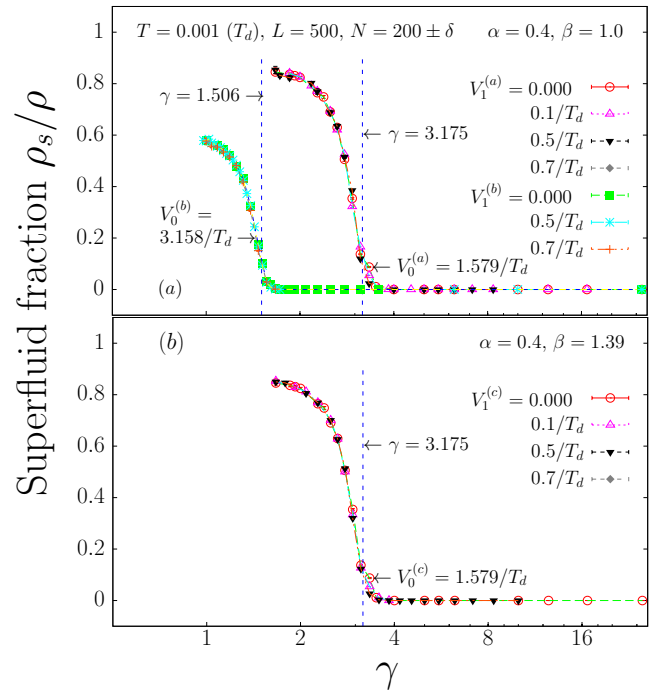}
\caption{(Color online) WA superfluid fraction $\rho_s/\rho$ versus the Lieb-Liniger parameter $\gamma$ for an interacting 
Bose gas in several realizations of the BCOL. The figure demonstrates primarily the WA numerical reproduction of the Sine-Gordon 
transition observed experimentally in Ref.\cite{Haller:2010}. The temperature of the system is $\widetilde{T}=0.001$, 
its length $L$ is such that $2L/\lambda=200$, and the number of particles $N$ is very close to 200 with a small error $\delta$ 
that is within the range $\pm 1$. Two primary OL depths $V_0^{(a)}=1.579/T_d$ and $V_0^{(b)}=3.158/T_d$ are considered with 
different associated depths $V_1^{(a)}$ and $V_1^{(b)}$ of the superimposed secondary OLs, respectively. Each of the primary 
OLs has exactly 200 sites with an occupancy of one particle per site in the SG regime. Upper frame ($a$) is for a QPOL with 
$\alpha=0.4$ and $\beta=1.0$. Open circles: $V_1^{(a)}=0.000$ (purely periodic OL); open up triangles: $0.1/T_d$; solid down 
triangles: $0.5/T_d$; and solid diamonds: $0.7/T_d$. Solid squares: $V_1^{(b)}=0.000$; stars: $0.5/T_d$; and crosses: $0.7/T_d$. 
Lower frame ($b$) is for a quasidisordered OL with $V_0^{(c)}=1.579/T_d$, the same $\alpha$, but a different $\beta=1.39$. The
labels for $V_1^{(c)}$ are the same as for $V_1^{(a)}$. The vertical dashed lines show the SG transition points $\gamma=3.175$ 
for $V_0^{(a)}=V_0^{(c)}=1.579/T_d$, and $1.506$ for $V_0^{(b)}=3.158/T_d$ which are very close to the ones found on the phase 
diagram Fig.3 of Haller \ea\ \cite{Haller:2010}. $V_0$, $V_1$, and $\widetilde{T}$ are in units of $T_d$ and $\gamma$ is unitless.}
\label{fig:plotEffectofDisorderSineGordonPhaseTransitionT0.001TdN200L500d0.400}
\end{figure}

\subsection{Chemical potential and interactions}

\hs In the simulations below, the number of particles $N$ for each $\gamma$ was fixed extremely close to 200 by a 
careful tuning of the chemical potential $\mu$ from a calibration curve, i.e., a plot of $N$ versus $\mu$ for each 
interaction strength $\gamma$. The error bars $\delta$ in $N$ were within $\pm 1.0$ and this error is unavoidable 
as one is dealing with a continuous-space Monte Carlo simulation. Fig.\ft\ref{fig:plot.mu.vs.gamma.several.A.and.B} 
displays a plot of the calibrated $\mu$ versus $\gamma$ for a number of BCOL realizations. The $\mu$ rises in 
general with $\gamma$ until it begins to stabilize in the strongly-interacting SG regime.

\section{Results}\label{sec:results}

\subsection{Sine-Gordon transition}\label{sec:SG-transition}

\hs An important result of this work is that the WA is able to reproduce the SG transition 
\cite{Haller:2010} in a primary 1D shallow periodic OL. It is found that the latter transition is 
robust against perturbations arising from the addition of a weaker secondary OL. The critical value 
$\gamma_c$ at which the transition occurs in a BCOL (\ref{eq:quasiperiodic-optical-lattice}) remains 
exactly the same as compared to a periodic OL with the same $V_0$, unaffected by the latter perturbations. 
This is the chief result of this paper and is demonstrated by the behavior of $\rho_s/\rho$ as a function 
of $\gamma$ for a Bose gas in various realizations of a 1D BCOL (\ref{eq:quasiperiodic-optical-lattice}). 
Fig.\ft\ref{fig:plotEffectofDisorderSineGordonPhaseTransitionT0.001TdN200L500d0.400} 
demonstrates this finding for two values of the primary depth $V_0^{(a)}$ and $V_0^{(b)}$ 
and different strengths of the associated secondary OL, $V_1^{(a)}$ and $V_1^{(b)}$, respectively.
Here, $\rho_s/\rho$ displays a steep decline towards the critical $\gamma_c$ beyond which it remains 
zero deep into the SG regime $\gamma\gg \gamma_c$. The value of $\gamma_c$ indicated by a vertical 
dashed line is obtained by fitting a function of the form $f(\gamma)\,=\,A(\gamma-\gamma_c)^s$ to the 
data of $\rho_s/\rho$ in a narrow range of $\gamma$ where $\rho_s/\rho$ comes close to zero. Here $A$, 
$\gamma_c$, and $s$ are fitting parameters. We shall return to this issue in a future publication 
where a comparison between WA numerical and experimental results is required in order to explore the 
validity of the SG theory \cite{Astrakharchik:2015}. The addition of a secondary OL does not alter the 
behavior of $\rho_s/\rho$ from the one observed for $V_1^{(a,b,c)}=0$, the purely periodic OL. For 
$V_0^{(c)}\,=\,1.579/T_d$, the behavior of $\rho_s/\rho$ in frame ($b$) with $\beta/\alpha=3.475$ is 
exactly the same as in frame ($a$) with $\beta/\alpha=2.5$ and the same $V_0^{(a)}\,=\,1.579/T_d$. This 
is a rather peculiar result showing that an increased quasidisorder [cf. Eq.(\ref{eq:delta-disorder}) 
with $\epsilon=\beta/\alpha$] in a shallow BCOL does not alter the behavior of $\rho_s/\rho$. The values 
of $\gamma_c$ at which the transitions occur for both values of $V_0$ are very close to the ones obtained 
from the phase diagram in Fig.~3 of Haller \ea\ \cite{Haller:2010}. This is again a striking demonstration 
of the fact that the WA is a powerful method enabling an accurate simulation of the behavior of lattice 
bosons in continuous space.

\hs The robustness of $\rho_s/\rho$ against the secondary-OL perturbations arises because the particles 
always seek the lowest energy states of the BCOL, i.e., those of the primary OLs. Further, in the 
strongly-interacting regime $\gamma>1$, the interactions override the effects introduced by $V_1$ as 
the behavior of bosons is chiefly dictated by the commensurate filling of the primary OL. This finding
is brought in line with that of Boeris \ea\ \cite{Boeris:2016}, who argued that in the limit of a shallow 
periodic potential the optical depth becomes subrelevant when the Mott transition is only controlled by 
interactions.

\begin{SCfigure*}
\includegraphics[width=8.5cm,bb=59 109 332 680,clip]{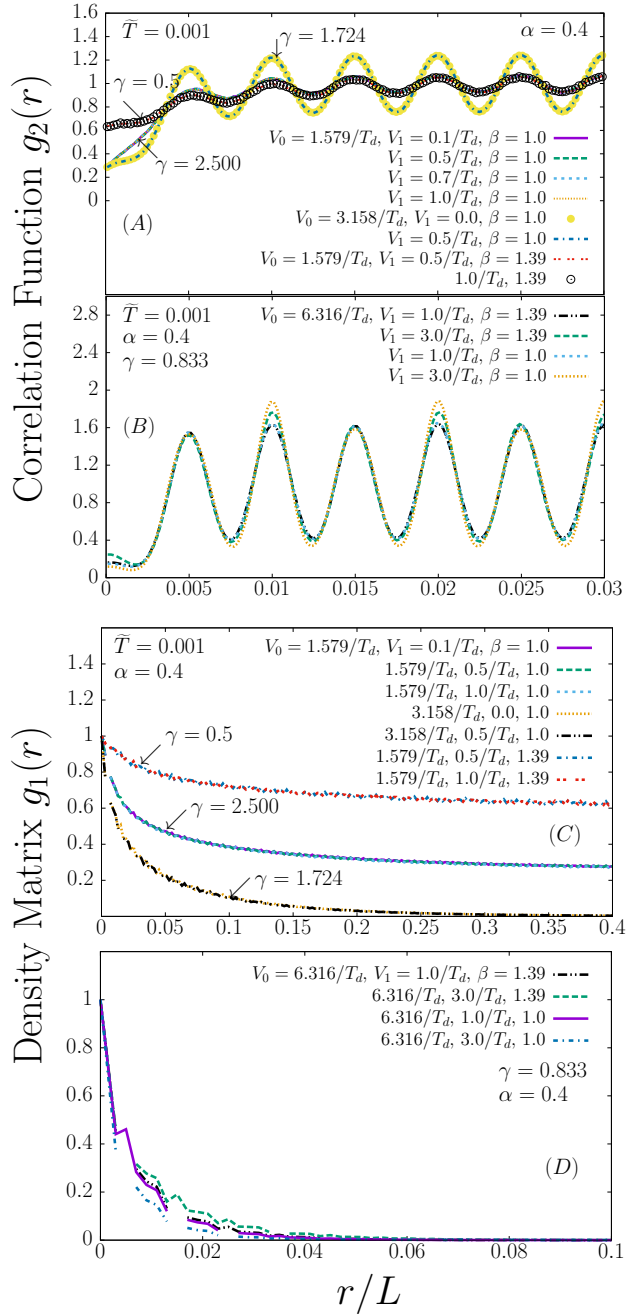}
\caption{(Color online) Effect of secondary optical lattice on the spatially-averaged correlation function $g_2(r)$ 
[Eq.(\ref{eq:g2r})] and the spatially-averaged density matrix $g_1(r)$ [Eq.(\ref{eq:obdm-homogeneous})] of the systems 
in Fig.\ft\ref{fig:plotEffectofDisorderSineGordonPhaseTransitionT0.001TdN200L500d0.400}. Here $r$ is the distance
between any pair of particles along the lattice and all frames share the same $x-$axis label. Frame ($A$) 
shows $g_2(r)$ for $\alpha=0.4$ at different interaction strengths and for various BCOL realizations. For 
$\gamma=2.500$, $V_0=1.579/T_d$, and $\beta=1$, the figure displays 
the cases for: $V_1=0.1/T_d$ (solid line); $0.5/T_d$ (long-dashed line); $0.7/T_d$ (short-dashed line); $1.0/T_d$ 
(fine-dotted line). (It is difficult to plot these lines so that all symbols can be distinguished from each other 
because they're exactly overlapping. The same applies to the rest of the plots.) For $\gamma=1.724$, $V_0=3.158/T_d$, 
and the same $\beta$, the figure shows the cases for $V_1=0.0$ (solid circles) and $0.5/T_d$ (dashed-dotted line). The 
last set of data is for $\gamma=0.5$, $V_0=1.579/T_d$ but with $\beta=1.39$ and: $V_1=0.5/T_d$ (double-dotted line) 
and $1.0/T_d$ (open circles). Frame ($B$) is as in ($A$); except for $V_0=6.316/T_d$, $\gamma=0.833$, and: $V_1=1.0/T_d$ 
with $\beta=1.39$ (dashed double-dotted line); $3.0/T_d$ with 1.39 (long-dashed line); $1.0/T_d$ with 1.0 (short-dashed 
line); and $3.0/T_d$ with 1.0 (fine-dotted line). Going over to $g_1(r)$, frame ($C$) displays it for $\gamma=2.500$, 
$V_0=1.579/T_d$, and: $V_1=0.1/T_d$ with $\beta=1.0$ (solid line); $0.5/T_d$ and 1.0 (dashed line); $1.0/T_d$ and 1.0 
(dotted line). The following set of data is for $\gamma=1.724$, $V_0=3.158/T_d$ and: $V_1=0.000$ with $\beta=1.0$ 
(fine-dotted line); $0.5/T_d$ with $\beta=1.0$ (dashed double-dotted line). Next comes $V_0=1.579/T_d$, $\gamma=0.5$, 
and $V_1=0.5/T_d$ with $\beta=1.39$ (dashed-dotted line); $1.0/T_d$ and 1.39 (double-dotted line). Finally, frame ($D$) 
is as in ($C$); except for $V_0=6.316/T_d$, $\gamma=0.833$, and: $V_1=1.0/T_d$ with $\beta=1.39$ (dashed double-dotted 
line); $3.0/T_d$ and 1.39 (dashed line); $1.0/T_d$ and 1.0 (solid line); $3.0/T_d$ and 1.0 (dashed-dotted line). The 
$r$ is in units of $L$, $\widetilde{T}$ is in units of $T_d$, and $\gamma$ is unitless.}
\label{fig:plotdensmcorrA1.579severalBasc2.000T0.001L500N200}
\end{SCfigure*}

\begin{figure}[t!]
\includegraphics[width=6.5cm,bb=67 395 538 683,clip]{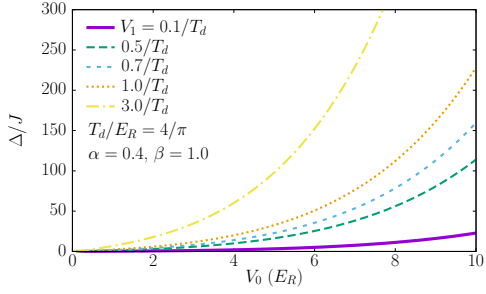}
\caption{(Color online) $\Delta/J$ [given by Eqs.(\ref{eq:delta-disorder}) and (\ref{eq:J-tunneling-amplitude})] 
as a function of the primary OL depth $V_0$ for various values of $V_1$: $0.1/T_d$ (solid line); $0.5/T_d$ 
(long-dashed line); $0.7/T_d$ (short-dashed line); $1.0/T_d$ (dotted line); and $3.0/T_d$ (dashed double-dotted line). 
The conversion factor from units of $T_d$ to $E_R$ is $T_d/E_R=4/\pi$, $V_0$ is in units of $E_R$, and $\Delta/J$
is unitless. }
\label{fig:plotdeltaandJ}
\end{figure}

\begin{table}
\caption{Ratio between the disorder strength $\Delta$ [Eq.(\ref{eq:delta-disorder})] and the tunneling
amplitude $J$ [Eq.(\ref{eq:J-tunneling-amplitude})] for various realizations of the BCOL. The table lists
from left to right the primary OL depth $V_0$, the secondary one $V_1$, the parameters determining the 
periodicity of the BCOL (\ref{eq:quasiperiodic-optical-lattice}), $\alpha$ and $\beta$, the parameters 
$\Delta$ and $J$, and then $\Delta/J$. The OL depths are in units of $T_d$, whereas $\Delta$ and $J$ are 
in units of $E_R$.}
\begin{tabular}{*7{@{\hspace{0.3cm}} c @{\hspace{0.2cm}}}} \\ \hline\hline
$V_0$ &  $V_1$ & $\alpha$ & $\beta$ & $\Delta$ & $J$ & $\Delta/J$ \\ 
($T_d$) & ($T_d$) & & & ($E_R$) & ($E_R$) & \\ \hline
1.579 &  0.1 &  0.4 &  1.0  &  0.0224  &  0.1804  & 0.1240 \\ 
      &  0.5 &      &       &  0.1119  &  0.1804  & 0.6200 \\
      &  0.7 &      &       &  0.1566  &  0.1804  & 0.8679 \\
      &  0.1 &      &  1.39 &  0.0432  &  0.1804  & 0.2395 \\
      &  0.5 &      &       &  0.2161  &  0.1804  & 1.1977 \\
      &  0.7 &      &       &  0.3026  &  0.1804  & 1.6768 \\
3.158 &  0.5 &      &  1.0  &  0.2349  &  0.1510  & 1.5553 \\
      &  0.7 &      &       &  0.3288  &  0.1510  & 2.1775 \\
6.316 &  1.0 &      &       &  0.7662  &  0.0886  & 8.6495 \\ 
      &  3.0 &      &       &  2.2987  &  0.0886  & 25.948 \\
      &  1.0 &      &  1.39 &  1.4805  &  0.0886  & 16.712 \\ 
      &  3.0 &      &       &  4.4414  &  0.0886  & 50.135 \\ \hline
\end{tabular}\label{table:deltaoverJ}
\end{table}

\subsection{Effect of secondary lattice}

\hs Figure \ref{fig:plotdensmcorrA1.579severalBasc2.000T0.001L500N200} shows the influence of $V_1$ on the 
spatial behavior of $g_1(r)$ and $g_2(r)$, particularly the amplitudes of the $g_2(r)-$oscillations, for
various realizations of the BCOL. A significant feature is that $g_2(r)$ reveals a spatial oscillatory
structure whose origin can be traced back to the primary OL. On the other hand, $g_1(r)$ in ($C$) displays a 
structure similar to $g_1(r)$ of the 1D homogeneous Bose gas in Fig.~\ref{fig:plot.obdm.vs.nr.and.sff.vs.T.two.figures}
without any spatial oscillations. 

\hs As $V_0$ is increased, so do the oscillatory amplitudes of $g_2(r)$ in ($B$) displaying its deep connection to 
the OL. An enhanced spatial decay rate of $g_1(r)$ at larger $V_0$ in ($C$) indicates a reduction of the superfluid 
fraction. The spatial frequency of the $g_2(r)-$oscillations neither changes with $V_0$ nor with $V_1$, and is largely 
governed by the lattice parameter of the primary OL. When $V_0$ and $V_1$ are small, e.g. $\sim 1.5/T_d$ in frames 
(A) and (C), a change of $V_1$ does not alter the spatial behavior. Remarkably, the secondary OL is practically not 
``seen" by the bosons in this case. The $g_1(r)$ and $g_2(r)$ for bosons in a QPOL ($\beta=1.0$) do not differ from 
those in a QDOL ($\beta=1.39$). However, when $V_0$ is increased to larger values $\sim 6/T_d$ as in ($B$) and ($D$), 
a change in the band structure of the BCOL via $V_1$ and $\beta$ begins to assert itself in $g_2(r)$ and $g_1(r)$. 
Further, at larger $V_0=6.316/T_d$, the change in the oscillatory amplitude of $g_2(r)$ with $V_1$ is more pronounced 
for a QPOL than for a QDOL indicating that an increased disorder does not necessary yield a stronger response to $V_1$. 
The same argument can be applied to $g_1(r)$ in frame (D). The secondary OL begins to influence the properties only 
in conjunction with a larger $V_0$ at which the BCOL begins to compete with the boson-boson interactions. Indeed, 
the difference in $\beta/\alpha$ introduces a difference between the band structures of the QPOL and the QDOL and 
therefore the corresponding $g_2(r)$.

\hs One can also explain the stronger response to $V_1$ by examining $\Delta/J$ for a noninteracting Bose 
gas in a BCOL. A plot of $\Delta/J$ [cf. Eqs.(\ref{eq:delta-disorder}) and (\ref{eq:J-tunneling-amplitude})] 
as a function of $V_0$, shows that $\Delta/J$ increases significantly at the larger $V_1$. Naturally, 
according to Eq.(\ref{eq:delta-disorder}), $\Delta$ rises with $\epsilon$ and $V_1$. Fig.\ft\ref{fig:plotdeltaandJ} 
demonstrates these facts for a number of $V_1$ values and for $\alpha=0.4$ and $\beta=1.0$ only. In addition, Table 
\ref{table:deltaoverJ} lists $\Delta/J$ for the various realizations of the BCOL in the present work. It can 
be seen for example that for $V_0=6.316/T_d$ and $\beta=1.39$, $\Delta/J$ begins to attain values that are much 
larger than for the rest of the BCOL parameters. As such, the properties of the system should begin to change 
significantly at this $V_0-$height. For the other OL depths and $\beta-$values, $\Delta/J$ has relatively 
small values of the order of $\sim 1$, and therefore, the secondary OL is unable to cause any changes in the 
properties.

\begin{figure}
\includegraphics[width=7.0cm,bb=172 212 440 701,clip]{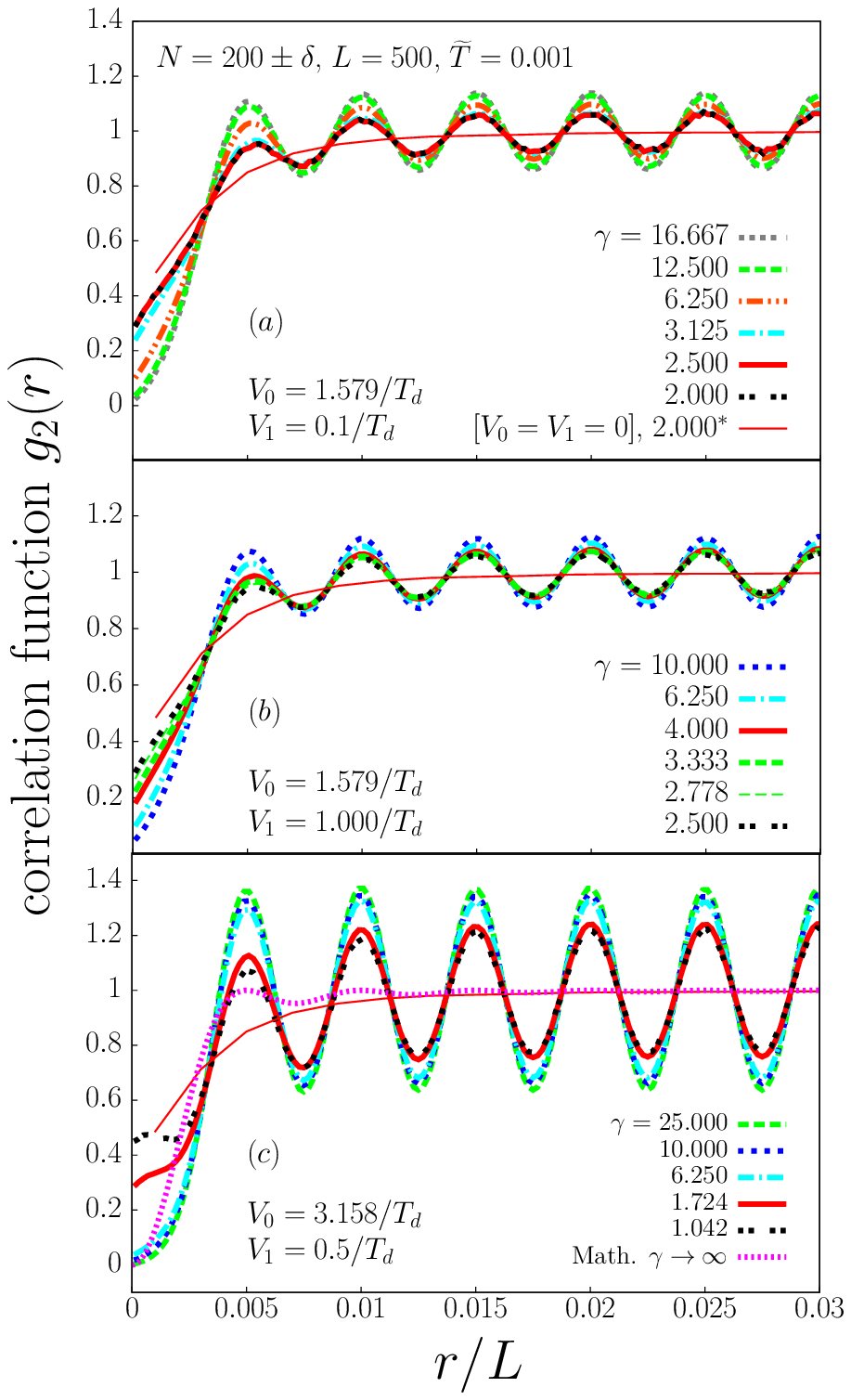}
\caption{(Color online) WA pair correlation function $g_2(r)$ [Eq.(\ref{eq:g2r})] at various interactions $\gamma$ 
for the systems of Fig.\ft\ref{fig:plotEffectofDisorderSineGordonPhaseTransitionT0.001TdN200L500d0.400}($a$) and 
three realizations of the BCOL with $\alpha=0.4$ and $\beta=1.0$. Frame ($a$) is for the BCOL of primary depth 
$V_0=1.579/T_d$ and a secondary depth $V_1=0.1/T_d$ at $\gamma=16.667$ (dotted line); 12.500 (dashed line); 
6.250 (dashed triple-dotted line); 3.125 (dashed-dotted line); 2.500 (thick solid line); and 2.000 (double-dotted 
line). For additional comparison, the thin solid line is exceptionally displayed for a homogeneous Bose gas without 
an OL [$V_0=V_1=0$] at $\gamma=2.000$ and the same parameters as in 
Fig.\ft\ref{fig:plotEffectofDisorderSineGordonPhaseTransitionT0.001TdN200L500d0.400}($a$). It is also displayed in
frames ($b$) and ($c$). Frame ($b$) is as in ($a$); but for $V_1=1.0/T_d$ and $\gamma=10.000$ (dotted line); 
6.250 (dashed-dotted line); 4.000 (thick-solid line); 3.333 (dashed line); 2.778 (thin-dashed line); 2.500 
(double dotted line). Frame ($c$) is for $V_0=3.158/T_d$, $V_1=0.5/T_d$, and: $\gamma=25.000$ (dashed line); 
10.000 (dotted line); 6.250 (dashed-dotted line); 1.724 (thick solid line); 1.042 (double-dotted line). The 
fine-dotted line shows the pair correlation function. The thin-dashed and thin-dashed-dotted lines show the
analytic results $g_2^{(per)}(z,1/\gamma)$ with $z=n\pi r$ [Eq.(\ref{eq:g2-per-1/gamma})] at $\gamma=10$ and 
$g_2^{(per)}(r,\sqrt{\gamma})$ [Eq.(\ref{eq:g2-per-sqrt-gamma})] at $\gamma=0.1$ from Ref.\cite{Sykes:2008}. The plot 
for these functions was generated by Mathematica${}^{\circledR}$. There is almost perfect commensurate filling 
with $N=200\pm\delta$ particles, where $\delta$ is a small error in the range $\pm 1$. $V_0$, $V_1$, and 
$\widetilde{T}$ are in units of $T_d$, and $\gamma$ is unitless.}
\label{fig:plotcorrGCANseveralAandBT0.001den0.400L500N200severalGammaStack}
\end{figure}

\subsection{Effect of interactions on the correlations}

\hs Fig.\ft\ref{fig:plotcorrGCANseveralAandBT0.001den0.400L500N200severalGammaStack} demonstrates the 
effect of interactions on $g_2(r)$ for all realizations of the BCOL. The qualitative pattern of these 
oscillations hardly changes with $\gamma$ (and $V_1$), even on going through the SG transition. 
Quantitatively, however, the amplitude of these oscillations rises in general in response to an increase 
of $\gamma$ (and $V_0$). [Compare frame $(c)$ with both $(a)$ and $(b)$.] The latter manifests an increase 
in the interplay between the interactions and the BCOL that in turn enhances the boson-boson 
(i.e., density-density $n_c(r)$) correlations. It must be emphasized that the rise in the amplitude of 
$g_2(r)-$oscillations is particularly significant beyond $\gamma_c$ reaching deep into the SG regime where 
the bosons are pinned. This is therefore indicative of a Mott insulator (MI) state which is not inert. An 
examination of the MGF $G(p=0,\tau)$ in Sec.~\ref{sec:MGF} and by making connections to $g_2(r)$, it is 
inferred that holes, arising from particle-hole (p$-$h) excitations, play an important role in the enhancement 
of $g_2(r)$.  

\hs On the one hand, the response of $g_2(r)$ to $\gamma$ can be further clarified by other arguments. 
First, it is known that $g_2(r)$ [Eq.(\ref{eq:g2r})] describes the probability of finding two particles 
at a separation $r$. Hence, as the bosons become more localized with increasing $\gamma$, the probability 
--i.e., the amplitude of $g_2(r)-$oscillations-- rises. Second, when the interactions are large, higher 
harmonics in the density operator $\widetilde{\rho}(x)$ from Ref.\cite{Giamarchi:2003} become excited that 
play a role in intensifying $g_2(r)$. Here, $\psi(x)=\sqrt{\widetilde{\rho}(x)} \exp(i\varphi(x))$ is a 
field operator with $\varphi(x)$ a phase and \cite{Giamarchi:2003,Roux:2008}

\begin{eqnarray}
&&\widetilde{\rho}(x)~=~\nonumber\\
&&\left\{n^2+\frac{1}{\pi^2}[\nabla\phi(x)]^2+n^2 \sum_{p>1} \cos(2\pi n x - 2 p \phi(x))\right\}^{1/2}
\nonumber\\ \label{eq:Giamarchi-rho(x)}
\end{eqnarray}

where $\phi(x)$ is a boson field operator and $n$ the average density. On the other hand, the rise in the
amplitude with $V_0$ can be related to the amplitude of the Wannier state in each OL well. For sufficiently
deep OLs, the Wannier state can be approximated by a harmonic oscillator ground state \cite{Boers:2007}

\begin{equation}
W(x)~=~\frac{k^{1/2}}{\pi^{1/4}}\left(\frac{V_0}{E_r}\right)^{1/8} 
\exp\left[-\frac{1}{2}\left(\frac{V_0}{E_r}\right)^{1/2} k^2 x^2\right] 
\label{eq:Wannier-state-Boers}
\end{equation}

where $k$ is the primary lattice wavevector. The amplitude of $W(x)$ rises with $V_0$ at each lattice site
and consequently the amplitude of $g_2(r)$. Moreover, it has been found \cite{Cherny:2009}, that correlations
arise from the interplay of quantum statisics, interactions, thermal, and quantum fluctuations, the last
of which can be related to the higher harmonics in Eq.(\ref{eq:Giamarchi-rho(x)}).

\subsection{Another test of the WA for $g_2(r)$}

\hs For the homogeneous Bose gas, we display in 
Fig.\ft\ref{fig:plotcorrGCANseveralAandBT0.001den0.400L500N200severalGammaStack}($c$) the analytic result for $g_2(r)$ 
perturbative in $1/\gamma$ \cite{Sykes:2008}

\begin{eqnarray}
&& g_2^{(per)}(z,1/\gamma)=1-\frac{\sin^2z}{z^2}-\frac{4}{\gamma}\frac{\sin^2z}{z^2}-\frac{2\pi}{
\gamma}\frac{\partial}{\partial z}\frac{\sin^2z}{z^2} + \nonumber\\
&&\frac{2}{\gamma}\frac{\partial}{\partial z}\left[
\frac{\sin^2z}{z^2}\int_{-1}^1 dt\,\sin(zt){\rm ln}\frac{1+t}{1-t}
\right]\label{eq:g2-per-1/gamma}
\end{eqnarray}

where $z=n\pi r$ and the one perturbative in $\gamma$

\begin{equation}
g_2^{(per)}(r,\sqrt{\gamma})=1-\sqrt{\gamma}\left[{\rm L}_{-1}\left(2\sqrt{\gamma}n r\right)-
{\rm I}_1\left(2\sqrt{\gamma}n r\right)\right],
\label{eq:g2-per-sqrt-gamma}
\end{equation}

where, ${\rm L_{-1}}(x)$ and ${\rm I}_1(x)$ are the Struve and Bessel functions, respectively. 
Eq.(\ref{eq:g2-per-1/gamma}) is plotted with $\gamma=10$ for the strongly, whereas (\ref{eq:g2-per-sqrt-gamma}) 
with $\gamma=0.1$ for the weakly interacting regime. The goal is to check the WA $g_2(r)$ [e.g. for $\gamma=2.000$]
without an OL against analytic calculations, and one can see that the WA result lies largely intermediate 
between these analytic results. This shows again that WA is reliable in calculating these properties. 

\subsection{Origin of oscillations}

\hs Despite the fact that (\ref{eq:g2-per-1/gamma}) diplays weak oscillations, these are substantially enhanced 
in the WA $g_2(r)$ by the addition of an OL. The WA $g_2(r)$ for the Bose gas without an OL does not 
show any oscillations whatsoever [e.g. for $\gamma=2.000$ and no OL in 
Fig.\ft\ref{fig:plotcorrGCANseveralAandBT0.001den0.400L500N200severalGammaStack}]. It is known 
that strongly-repulsive bosons in 1D without an OL arrange themselves in a periodic structure as if they were 
ordered inside an OL, but still the $g_2(r)$ reveals no spatial oscillations. Fact is, that the OL introduces
a different energy level structure that reshapes the wavefunction into a spatially oscillatory form.

\hs Let us here only mention briefly the results of the appendices and the reader is 
referred to the them for more information. In Appendix \ref{app:norm-corr-function-density}, the SAPCF $h_2(r)$ 
[Eq.(\ref{eq:normcorrh2r})] and the SA-OBDM $h_1(r)$ [Eq.(\ref{eq:obdm-inhomogeneous})] are displayed in 
Figs.\ft\ref{fig:plotnormdensmnormcorrA1.579severalBasc2.000T0.001L500N200} 
and \ref{fig:plotnormcorrGCANseveralAandBT0.001den0.400L500N200severalGammaStack} (for the same systems 
of Figs.\ft\ref{fig:plotdensmcorrA1.579severalBasc2.000T0.001L500N200} and 
\ref{fig:plotcorrGCANseveralAandBT0.001den0.400L500N200severalGammaStack}, respectively). The goal is to 
reveal the effects introduced into the correlations when instead of $n_0^2$ and $n_0$, the $g_2(r)$ and 
$g_1(r)$ are normalized by $\rho_c(r)$ and $\rho_{c,\frac{1}{2}}(r)$ [Eq.(\ref{eq:spatially-varying-density-norm}) 
and (\ref{eq:rho-half})], respectively. [That is $g_1(r)\rightarrow h_1(r)=g_1(r)n_0/\rho_{c,\frac{1}{2}}(r)$ and 
$g_2(r)\rightarrow h_2(r)=g_2(r)n_0^2/\rho_c(r)$.] The result is that in general --for a shallow BCOL-- $h_2(r)$ 
displays oscillations with a (much) smaller amplitude than $g_2(r)$. In fact for a shallow BCOL, $h_2(r)$ is seen to 
approach the behavior of $g_2(r)$ for a homogeneous Bose gas! It can therefore be concluded, that in this case the 
origin of the $g_2(r)-$oscillations is the oscillations in the spatial density whose structure is determined by the 
OL. In Appendix \ref{app:norm-corr-function-density}, except for small oscillations, 
$h_1(r)$ shows generally the same qualitative behavior as $g_1(r)$.

\begin{figure}[t!]
\includegraphics[width=8.5cm,bb=70 373 532 682,clip]{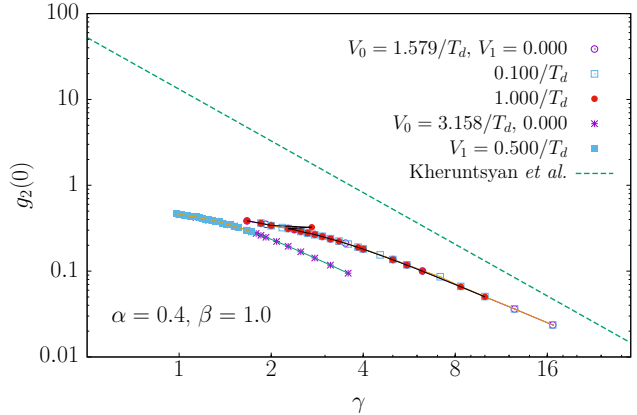}
\caption{(Color online) Spatially-averaged WA local pair correlation function $g_2(r=0)$ [Eq.(\ref{eq:g2r})] as 
a function of $\gamma$ for the systems of Fig.\ft\ref{fig:plotcorrGCANseveralAandBT0.001den0.400L500N200severalGammaStack} 
in comparison to $g^{(2)}(\gamma)$ of Kheruntysan \ea\ \cite{Kheruntsyan:2003} (dashed line) for the homogeneous Bose
gas in the strongly interacting regime $\gamma>1$. Here, the QPOL is with $\alpha=0.4$ and $\beta=1.0$. 
The figure displays $g_2(0)$ for $V_0=1.579/T_d$ with $V_1=0$ (open circles); with $V_1=0.100/T_d$ (open 
squares); and with $V_1=1.000/T_d$ (solid circles). The stars are for $V_0=3.158/T_d$ with $V_1=0.000$ and 
with $V_1=0.5/T_d$ (solid squares).}\label{fig:compareAnalyticalcorr0withWAPIMC}
\end{figure}

\subsection{Interaction energy and local $g_2(0$)}

\hs According to Astrakharchik \ea\ \cite{Astrakharchik:2005}, the total interaction energy $E_{int}$ can 
be computed from $g_2(r)$ at $r=0$ using

\begin{equation}
\frac{E_{int}}{\langle N\rangle T_d}\,=\,\frac{1}{2} g_{1D} n g_2(0).
\end{equation}

Now $E_{int}$ goes to zero as $\gamma$ is increased to large values beyond $\gamma_c$ because $g_2(0)\rightarrow 0$. 
This can be seen in Fig.\ft\ref{fig:plotcorrGCANseveralAandBT0.001den0.400L500N200severalGammaStack} for 
$\gamma>6.250$ in all frames and is a stark indication to fermionization of the bosons \cite{Kheruntsyan:2003} 
and demonstrates perfect antibunching \cite{Sykes:2008}. In frame ($c$), fermionization is reached at $\gamma=10$ 
whereas in frames ($a$) and ($b$) it requires $\gamma>10$. Therefore, a larger $V_0$ aids the fermionization 
process of bosons since there is an enhanced effective interaction arising from the interplay between the BCOL 
and the repulsive forces that reduces the value of $\gamma$ required for fermionization. The added secondary OL 
does not play a role in this regard.

\hs In Fig.\ft\ref{fig:compareAnalyticalcorr0withWAPIMC}, $g_2(0)$ is plotted as a function of
$\gamma$ for the two BCOLs with $V_0=1.579/T_d$ and $3.158/T_d$ and some values of $V_1$. For 
comparison, the analytical result for the homogeneous Bose gas of Ref.\cite{Kheruntsyan:2003},

\begin{equation}
g^{(2)}(\gamma)=\frac{4}{3} \frac{\pi^2}{\gamma^2} \left[1+\frac{t^2}{4\pi^2}\right],
\end{equation}

with $t=T/T_d$ is displayed by the linear dashed line. It is immediately observed that, whereas 
on the one hand the introduction of an OL to a homogeneous Bose gas reduces $E_{int}$ via a 
significant reduction of $g_2(0)$, the addition of a secondary weaker OL does not play much of 
a role in changing the behavior of $g_2(0)$ for the same $V_0$. The interaction in the system is 
therefore not influenced by a perturbation of the primary OL. The effect of $\gamma$ on $g_1(r)$ 
shall be explored in a future publication in connection to the change of its behavior across the 
SG transition. 

\hs In Appendix \ref{app:local-h2(0)}, $h_2(0)$ vs $\gamma$ is displayed in 
Fig.\ft\ref{fig:compareAnalyticalnormcorr0withWAPIMC} for the same systems of Fig.\ft\ref{fig:compareAnalyticalcorr0withWAPIMC}. 
Peculiarly, $h_2(0)$ shows qualitatively the same behavior of $g_2(0)$.

\begin{figure}[t!]
\includegraphics[width=7.0cm,bb=163 61 446 655,clip]{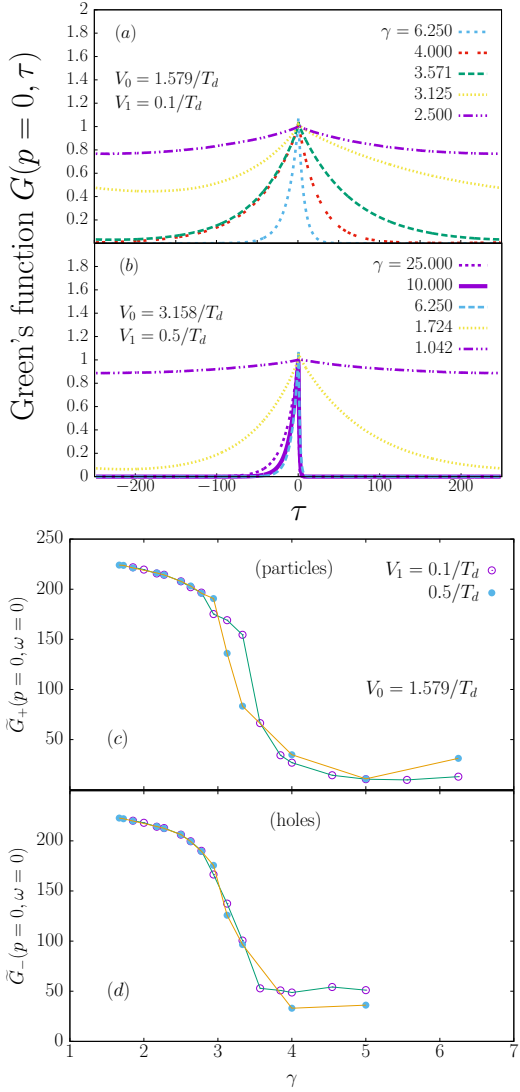}
\caption{(Color online) WA Matsubara Green's function $G(p=0,\tau)$ and weights of particle and hole (p-h) 
excitations $\widetilde{G}_+(p=0,\omega=0)$ and $\widetilde{G}_-(p=0,\omega=0)$ [Eq.(\ref{eq:FTG+G-})], 
respectively, for the systems of 
Fig.\ft\ref{fig:plotEffectofDisorderSineGordonPhaseTransitionT0.001TdN200L500d0.400}. Upper figure displays 
$G$ as a function of $\tau$ for some values of $\gamma$, and the lower $\widetilde{G}_\pm$ as a function of 
$\gamma$ --all for different values of $V_1$. Frame ($a$) displays $G$ for $V_0=1.579/T_d$ and $V_1=0.100/T_d$ with: 
$\gamma=6.250$ (dotted line); 4.000 (double-dotted line); 3.571 (dashed line); 3.125 (fine-dotted line); 2.500 
(dashed triple-dotted line). Frame ($b$) is as in ($a$); but with $V_0=3.158/T_d$, $V_1=0.5/T_d$ and: $\gamma=25.000$ 
(dotted line); 10.000 (solid line); 6.250 (dashed line); 1.724 (fine-dotted line); 1.042 (dashed triple-dotted line). 
Frame ($c$) shows $\widetilde{G}_+$ for $V_0=1.579/T_d$ and $V_1=0.1/T_d$ (open circles); $0.5/T_d$ (solid circles). 
Frame ($d$) is as in ($c$) with the same labels; but for $\widetilde{G}_-$. $T$ and $\mu$ are in units of $T_d$ and
$\gamma$ is unitless.}
\label{fig:plotGGGCANA1.579severalBT0.001den0.400L500N200severalGammaStack}
\end{figure}

\subsection{Matsubara Green's function}\label{sec:MGF}

\hs The MGF $G(p=0,\tau)$ displays (1) the presence of a gas of particle-hole (p-h) pairs revealing that 
the insulating state obtained following the SG transition is not inert similar to an MI \cite{Natu:2013}
and (2) that the weight of p-h excitations declines with increasing $\gamma$. This is shown in 
Fig.\ft\ref{fig:plotGGGCANA1.579severalBT0.001den0.400L500N200severalGammaStack}
frames ($a$) and ($b$), where $G(p=0,\tau)$ decays at a faster rate with $|\tau|$ as $\gamma$
is increased for any realization of the BCOL (we show here only two cases). In the limit when 
$\tau\rightarrow \pm\infty$, one can approximate $G(p=0,\tau)$ by \cite{Lewenstein:2012}

\begin{equation}
G(p=0,\tau) \sim
\left\{\begin{array}{l @{\quad;\quad} r}
e^{-E_P t} & \tau\rightarrow+\infty \\
e^{+E_H t} & \tau\rightarrow-\infty
\end{array}\right.
\end{equation}

where $E_P$ and $E_H$ are the single-particle and single-hole excitation energies, and $\tau=it$ 
is imaginary time. $G(p=0,\tau)$ reveals that for larger $\gamma$ and $\tau$ higher values of $E_P$ 
and $E_H$ are required to induce p-h excitations. The life-time of the p-h excitations $\sim \hbar/E_P$ 
or $\sim \hbar/E_H$, respectively, declines therefore with increasing $\gamma$ as well. This demonstrates 
that because of the high repulsion the bosons become locked in their positions unable to move unless 
excited by a strong external perturbation. But even after they are excited, they are deexcited after a 
short time as they return to the OL. Moreover, $G(p=0,\tau)$ is asymmetric about the origin $\tau=0$, and 
particularly beyond the SG transition point $\gamma_c$, $G(p=0,\tau)$ displays higher probability for hole 
excitations than particles.

\hs One can obtain a weight for the particle (hole) excitations from an integration of the area 
under $G(p=0,\tau)$ in the range $[0,\tau_{max}]$ ($[-\tau_{max},0]$), where $\tau_{max}=250$ is 
the maximum value of $\tau$. In fact, an integration from $-\tau_{max}$ to $+\tau_{max}$ would be 
approximately equivalent to the Fourier transform of $G(p=0,\tau)$ at a frequency of excitation 
$\omega=0$. We therefore consider for the present purpose

\begin{equation}
\widetilde{G}(p=0,\omega)\,\sim\,\int_{-\tau_{max}}^{+\tau_{max}} G(p=0,\tau) e^{-i\omega \tau} d\tau.
\label{eq:FTGreen}
\end{equation}

and separate it into two terms 

\begin{equation}
\widetilde{G}(p=0,\omega)\,=\,\widetilde{G}_+(p=0,\omega)\,+\,\widetilde{G}_-(p=0,\omega),
\label{eq:separatedG}
\end{equation}

where 

\begin{eqnarray}
\widetilde{G}_+(p=0,\omega)\,&\sim&\,\int_0^{\tau_{max}} G(p=0,\tau) e^{-i\omega\tau} d\tau,\nonumber\\
\widetilde{G}_-(p=0,\omega)\,&\sim&\,\int_{-\tau_{max}}^{0} G(p=0,\tau) e^{-i\omega\tau} d\tau. \nonumber\\
\label{eq:FTG+G-}
\end{eqnarray}

One can then define $\widetilde{G}_+(p=0,\omega=0)$ and $\widetilde{G}_-(p=0,\omega=0)$ as 
the weights of particle and hole excitations, respectively. $\omega=0$ is chosen because 
there is no excitation agent in our simulations.

\begin{table}[t!]
\caption{Standard deviation of the BCOL $\delta V$ [Eq.(\ref{eq:deltaV})] and $\Delta/J$ [cf. 
Eqs.(\ref{eq:delta-disorder}) and (\ref{eq:J-tunneling-amplitude})] for various realizations 
of the BCOL (\ref{eq:quasiperiodic-optical-lattice}). From left to right the table lists the 
primary OL depth $V_0$, the secondary depth $V_1$, the standard deviation $\delta V$ for 
$\alpha=0.4$ and $\beta=1.39$, $\delta V$ for $\widetilde{\alpha}=0.401248...$ and 
$\widetilde{\beta}=1.389964...$, $\Delta/J$ for ($\alpha$,$\beta$), and finally $\Delta/J$ 
for ($\widetilde{\alpha}$,$\widetilde{\beta}$), respectively. $V_0$, $V_1$, $\delta V$ are 
in units of ($T_d$), and $\Delta/J$ is unitless.}
\begin{tabular}{*6{|@{\hspace{0.2cm}} c @{\hspace{0.2cm}}}|}\hline
$V_0$   &  $V_1$   & $\delta V$ ($\alpha$,$\beta$) & $\delta V$ ($\widetilde{\alpha}$,$\widetilde{\beta}$)  & $\Delta/J$ & $\Delta/J$ \\ 
($T_d$) &  ($T_d$) & ($T_d$) & ($T_d$) & ($\alpha$,$\beta$) & ($\widetilde{\alpha}$,$\widetilde{\beta}$) \\ \hline\hline
1.579         & 0.1 & 0.5594 & 0.5595 & 0.2395 & 0.2380 \\
              & 0.5 & 0.5856 & 0.5857 & 1.1977 & 1.1900 \\
              & 0.7 & 0.6107 & 0.6108 & 1.6768 & 1.6600 \\
6.316         & 1.0 & 2.2609 & 2.2613 & 16.712 & 16.6032 \\
              & 3.0 & 2.4721 & 2.4726 & 50.138 & 49.8094 \\ \hline
\end{tabular}\label{tab:disorder-strengths-rational-irrational}
\end{table}

\hs Frames ($c$) and ($d$) display $\widetilde{G}_+(0,0)$ and $\widetilde{G}_-(0,0)$ as functions of 
$\gamma$, for $V_0=1.579/T_d$ and some values of $V_1$. Obviously, $\widetilde{G}_\pm$ decline with 
increasing $\gamma$ and, moreover, both frames show an apparent change in the curvature of $\widetilde{G}_\pm$ 
close to the critical $\gamma_c$ at which the SG transition occurs. Therefore, the change in the sign 
of $\partial^2 \widetilde{G}_\pm(0,0)/\partial\gamma^2$ is a signal for the SG transition, and 
$\partial^2 \widetilde{G}_\pm(0,0)/\partial\gamma^2=0$ near $\gamma=\gamma_c$.

\section{Rational vs irrational $\lambda_1/\lambda_2$}\label{sec:rational-vs-irrational}

\hs Whereas the ratio $\beta/\alpha\,=\,\lambda_1/\lambda_2$ is clearly rational for both pairs of 
$(\alpha,\beta)=(0.4,1.0)$ and $(0.4,1.39)$, it could be argued that had one used for example the 
irrational values $\widetilde{\alpha}\,=\,\sqrt{0.161}\,=\,0.401248\cdots$ which is very close to 0.4, or 
$\widetilde{\beta}\,=\,\sqrt{1.9320}\,=\,1.389964\cdots$ close to 1.39, the same results would be gotten
as for $(0.4,1.39)$. Within this context, Table \ref{tab:disorder-strengths-rational-irrational} lists 
$\delta V$ [Eq.(\ref{eq:deltaV})] and $\Delta/J$ [Eqs.(\ref{eq:delta-disorder}) and (\ref{eq:J-tunneling-amplitude})] 
for various realizations of the BCOL at ($\widetilde{\alpha}$,$\widetilde{\beta}$) and the rational 
approximation ($\alpha$,$\beta$). It can be seen that the results of $\delta V$ and $\Delta/J$ for 
($\alpha$,$\beta$) differ only slightly from that for ($\widetilde{\alpha}$,$\widetilde{\beta}$) by an 
order of magnitude $\stackrel{<}{\sim} 1\%$. In that sense, there should not be much difference in the results because 
of rational and irrational $\lambda_1/\lambda_2$.

\section{Conclusion}\label{sec:conclusion}

\hs In summary then, the properties of bosons in a shallow 1D bichromatic optical lattice (BCOL), 
with a rational ratio $\lambda_1/\lambda_2$ of its wavelengths, have been examined by scanning the 
system along a range of the Lieb-Liniger parameter $\gamma$ in the regime of the Sine-Gordon (SG) 
transition \cite{Haller:2010} at commensurate filling of the primary OL.

\hs The chief result is that this transition in a primary OL of depth $V_0\sim 1.5/T_d$ is robust 
against the perturbations by a secondary OL of depth $V_1<V_0$. The critical interaction strength 
$\gamma_c$ at which the transition occurs remains the same as for the periodic OL. The corresponding 
behavior of the superfluid fraction $\rho_s/\rho$ vs $\gamma$ reveals absolutely no response to changes 
in $V_1$; this is quite the same for other properties such as the correlation function $g_2(r)$ and the 
one-body density matrix $g_1(r)$. In contrast, changes in the latter observables arise with $V_1$ for 
larger values of $V_0\sim 6/T_d$ at which the lattice-band structure of the primary OL begins to be 
influenced by the secondary OL.

\hs However, for $V_0\sim 1.5/T_d$ the properties are significantly affected by changes in $\gamma$.
In this regard, $g_2(r)$ demonstrates an oscillatory structure whose amplitude rises with $\gamma$ and 
$V_0$, manifestating an increase in the interplay between BCOL and interactions that yield the excitation of
higher harmonics in the density operator. The origin of these oscillations lies chiefly in the primary 
OL. The latter changes are particularly significant beyond $\gamma_c$ reaching deep into the SG regime 
signalling the presence of a non-inert Mott insulator in the form of a particle-hole gas of bosons. At 
very large $\gamma$ beyond the critical $\gamma_c$, $g_2(0)\rightarrow 0$ signals fermionization and 
entrance to the Tonks-Girardeau regime. The fermionization process is aided by the primary OL and is 
unaffected by the secondary OL. Since $g_2(r)$ has been measured experimentally \cite{Kinoshita:2005} 
over a broad range of the coupling strengths, our work should then motivate a future experimental 
measurement of $g_2(0)$ in a BCOL.

\hs Although a division of $g_2(r)n_0^2$ and $g_1(r)n_0$ by the convoluted densities
$\rho_c(r)$ and $\rho_{c,\frac{1}{2}}(r)$, respectively, yields qualitatively the same results as 
far as the sensitivity to changes in the BCOL is concerned, the response of $h_2(r)=g_2(r)n_0^2/\rho_c(r)$ 
to changes in $V_0$ and $\gamma$ is in general significantly weakened. 
Hence, $g_2(r)$ is more adequate for obtaining stronger signals of changes. The $g_2(r)$ displays oscillations 
arising from oscillations in the spatial density that account for the same behavior in the density-density
correlations.

\hs The MGF $G(p=0,\tau)$ displays a declining weight for the particle-hole excitations with rising 
$\gamma$ attributed to an increased localization of the bosons. Moreover, it has been found that the 
system favors hole excitations at strong interactions leading us to conclude that deep in the SG regime 
holes play a chief role in the response of $g_1(r)$ and $g_2(r)$ to $\gamma$. In addition, a change 
in the curvature of the Fourier transform of the MGF $G(p=0,\omega=0)$ as a function of $\gamma$ 
for either particles or holes, is a signal for the SG transition.

\hs It has been particularly argued, that there shouldn't be much difference between the results 
obtained by using an irrational $\lambda_1/\lambda_2$ and a rational approximation to the same 
$\lambda_1/\lambda_2$ basing on the fact that the disorder strength of a BCOL changes only by small 
percentages if one uses a rational approximation to, instead of an irrational, $\lambda_1/\lambda_2$.

\hs Finally, it must be emphasized that the WA applied here has been tested for accuracy. By making 
comparions with previous results, it has been shown that the WA code works correctly and is therefore 
reliable. For example, our results for $g_1(r)$ are in very good agreement with those computed 
by Astrakharchik and Giorgini \cite{Astrakharchik:2005}. Moreover, the superfluid behavior with temperature 
matches exactly an analytical calculation from Ref.\cite{DelMaestro:2010}. Most importantly, the WA code 
was able to reproduce the SG transition observed earlier by Haller \ea\ \cite{Haller:2010}. The current 
results are also in line with Edwards \ea\ \cite{Edwards:2008} who demonstrated that the effects of weak 
perturbations to a primary 1D OL disappear as interactions get stronger and we {\it are} in the strongly
interacting regime. During the preparation of this paper, we learnt of an investigation bearing similarities
to ours by Bo$\acute{e}$ris \ea\ \cite{Boeris:2016} where, among other issues, the WA has been applied 
to reproduce the SG transition, except that they only use a periodic OL with a different procedure than 
ours here.

\acknowledgements

\hs Interesting and enlightning discussions with Sebastiano Pilati (ICTP, Trieste, Italy) 
are gratefully acknowledged and particulary for suggesting the topic on the Sine-Gordon transition.
Additional thanks go to Nikolay Prokofiev (UMASS, Amherst, USA) for interesting discussions
and for providing the worm algorithm code. The author wishes to thank the Abdus-Salam International
Center for Theoretical Physics (ICTP) in Trieste, Italy, and the Max Planck Institute for Physics of 
Complex Systems (MPIPKS) in Dresden Germany, both for providing him access to their excellent 
computational cluster and for a hospitable stay during scientific visits in which this work was 
undertaken. We are grateful to G. Astrakharchik for providing us with their data displayed
in Fig.~\ref{fig:plot.obdm.vs.nr.and.sff.vs.T.two.figures}. We thank William J. Mullin (UMASS, 
Amherst, USA) for a critical reading of the manuscript.This work has been carried out during 
the sabbatical leave granted to the author Asaad R. Sakhel from Al-Balqa Applied University (BAU) 
during academic year 2014/2015.

\appendix
\section{Worm Algorithm}\label{app:wpimc}

\hs Specifically, a WA code is applied which has been written by Nikolay Prokofev \cite{Prokofev:2011} 
originally for the simulation of 1D soft-core bosons without any trapping potential. A 3D version of this 
code, although for $^4$He on graphite, has been described earlier \cite{Boninsegni:2006}. WA is a path-integral 
Monte Carlo technique whose configurational space is extended to include what one calls ``worms". In 
this method, particles are described by Feynman's path-integral formulation as closed trajectories in 
space-time (diagonal configurations). Each trajectory is a closed chain of ``beads" and each ``bead" is 
a pair of positions of the same particle separated by a time $\tau$ along the imaginary-time axis in 
space-time. In a closed trajectory, the initial and final positions of the path are the same along the
space axis. The method considers imaginary ``time" as inverse temperature \cite{Mahan:1990}, i.e., 
$it/\hbar\leftrightarrow 1/(k_B T)$, where $\hbar$ is Planck's constant, $k_B$ Boltzmann's 
constant, and $T$ the temperature. The configuration of the system is updated by adding or removing 
open trajectories in space-time called ``worms" (off-diagonal configurations) which are paths in space-time 
whose initial and final positions along the space axis are not the same. Therefore, the configuration 
of the system is divided into two sectors: the $Z-$sector containing all the closed trajectories and the 
$G-$sector containing one open trajectory, or worm. The code is designed to perform various updates on 
these worms via certain acceptance and rejection probabilities that are carefully weighted functions of, 
and including, the factors $\exp[-(\mathbf{r}_\ell\,-\,\mathbf{r}_{\ell-1})^2/(4\lambda\tau)]$ and 
$\exp[-\tau V(\mathbf{r}_\ell)]$. Here $\lambda\,=\,\hbar^2/(2m)$ with $m$ the mass of a boson, 
$\tau\rightarrow 1/(M k_B T)$ is the imaginary time step with $M$ the total number of beads constituting 
the particle's trajectory along the imaginary time axis, $\mathbf{r}_\ell$ is the position of the $\ell^{th}$ 
bead along the space axis, and $V(\mathbf{r}_\ell)$ is the potential energy of bead $\ell$. The above 
factors included in a worm-update probability are integrated along the time axis over the length of the 
worm being updated. The beginning and end of the worm are termed MASHA and IRA, respectively, where MASHA 
precedes IRA along the imaginary time axis. There are various types of updates: (1) INSERT or REMOVE in 
which a worm is created and added to, or annihilated from, the configuration, respectively; (2) MOVE 
FORWARD or MOVE BACKWARD that change the length of the worm in space-time along the imaginary time axis; 
(3) RECONNECT in which a worm is connected to a closed trajectory after opening it resulting in an exchange 
of particles (permutation); (4) CUT or GLUE where a closed trajectory is cut open by removing, or an open
trajectory is closed by adding, a short chain of beads, respectively. All these updates are described in 
more detail by the inventors of the technique \cite{Boninsegni:2006}.

\begin{SCfigure*}
\includegraphics[width=8.5cm,bb=60 45 336 681,clip]{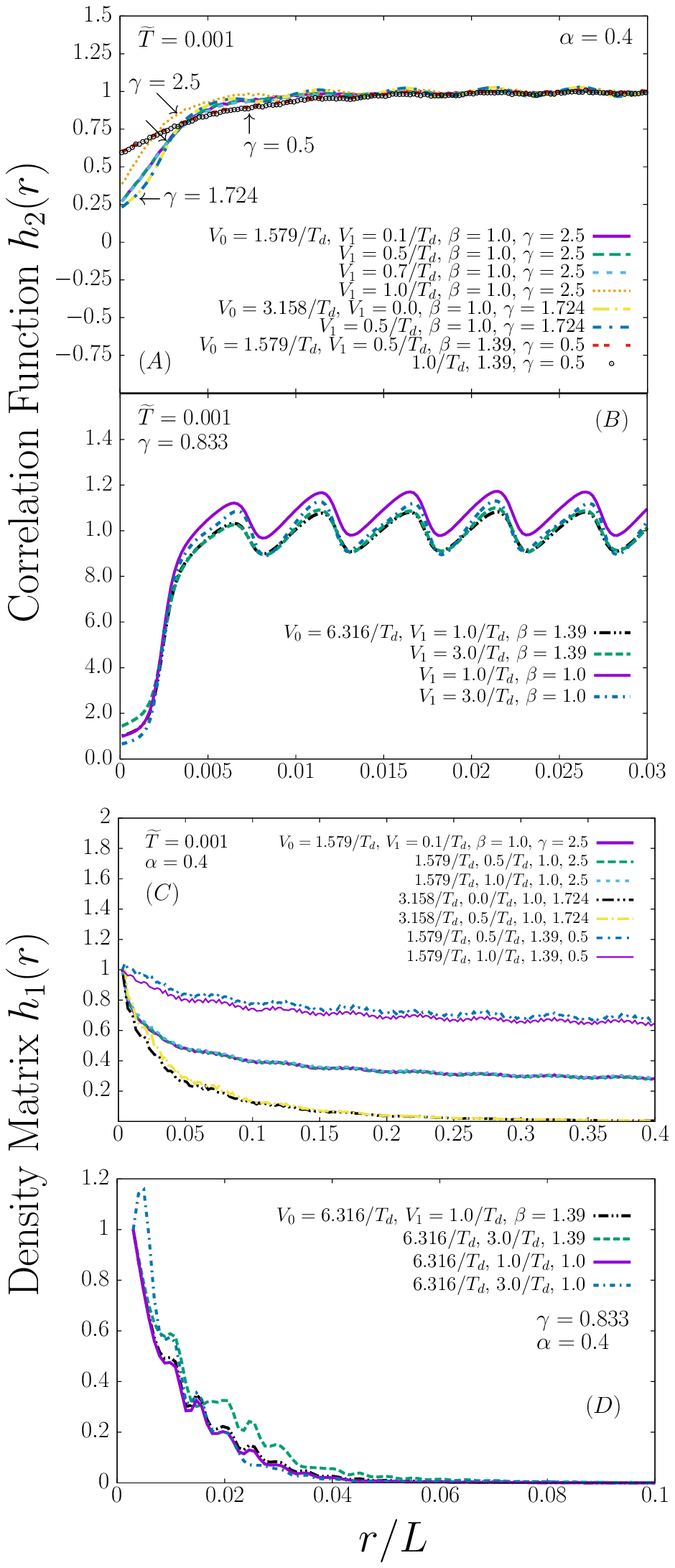}
\caption{(Color online) As in Fig.\ft\ref{fig:plotcorrGCANseveralAandBT0.001den0.400L500N200severalGammaStack}; but 
for $h_2(r)$. Upper frame ($A$): at $\gamma=2.5$ and $V_0=1.579/T_d$ the figure displays the results for: $V_1=0.1/T_d$, 
$\beta=1.0$ (thick solid line); 0.5, 1.0 (thick-dashed line); 0.7, 1.0 (short-dashed line); 1.0, 1.0 (fine-dotted line). 
Then at $\gamma=1.724$ and $V_0=3.158/T_d$ the results for: 0.0, 1.0 (long-dashed dotted line); 0.5, 1.0 (short-dashed dotted line). 
At $\gamma=0.5$ and $V_0=1.579/T_d$ again the results are shown for: $V_1=0.5/T_d$, $\beta=1.39$ (double-dotted dotted line); 
1.0, 1.39 (open circles). Upper frame ($B$) is for $V_0=6.316/T_d$, $\gamma=0.833$, and: $V_1=1.0/T_d$, $\beta=1.39$ 
(long-dashed double-dotted line); $3.0/T_d$, 1.39 (thick-dashed line); $1.0/T_d$, 1.0 (thin-dashed line); $3.0/T_d$, 1.0 
(dashed-dotted line). Lower frame ($C$) is for the same system in ($A$); but for $h_1(r)$ with $V_0=1.579/T_d$ and: $V_1=0.1/T_d$, 
$\beta=1.0$, $\gamma=2.5$ (thick solid line); $0.5/T_d$, 1.0, 2.5 (thick-dashed line); $1.0/T_d$, 1.0, 2.5 (thin-dashed line).
Then one has for $V_0=3.158/T_d$: $0.0/T_d$, 1.0, 1.724 (long-dashed double-dotted line); $0.5/T_d$, 1.0, 1.724
(dashed-dotted line). Next one has for $V_0=1.579/T_d$: $0.5/T_d$, 1.39, 0.5 (short-dashed dotted line);
$1.0/T_d$, 1.39, 0.5 (thin solid line). Finally frame ($D$) is for the same system in ($B$); but for $h_1(r)$:
dashed double-dotted line: $V_1=1.0/T_d$, $\beta=1.39$; dashed line: $3.0/T_d$, 1.0; thick solid line: $1.0/T_d$,
1.0; dashed-dotted line: $3.0/T_d$, 1.0.}
\label{fig:plotnormdensmnormcorrA1.579severalBasc2.000T0.001L500N200}
\end{SCfigure*}

\begin{figure}
\includegraphics[width=8.5cm,bb=170 216 439 654,clip]{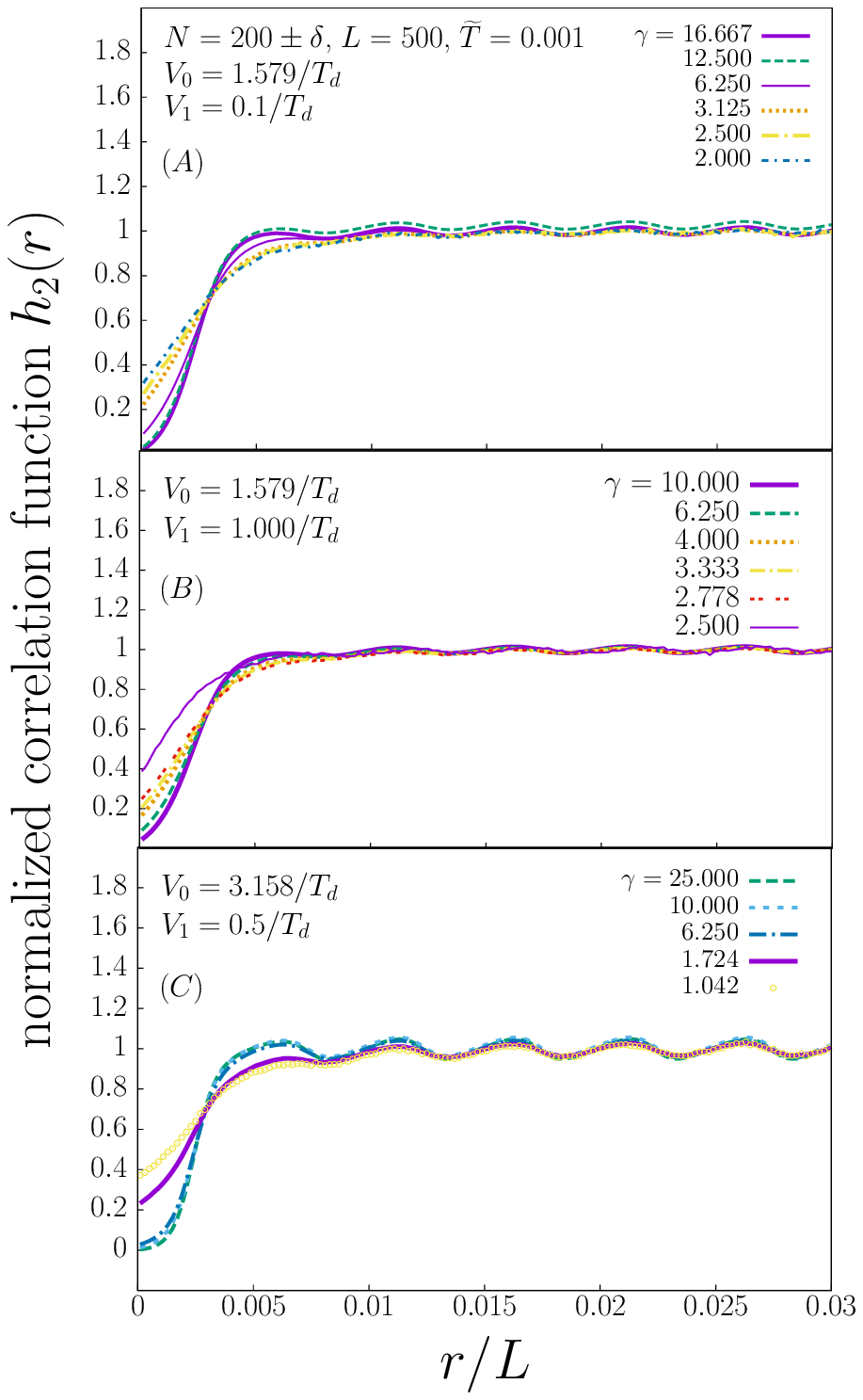}
\caption{As in Fig.\ft\ref{fig:plotcorrGCANseveralAandBT0.001den0.400L500N200severalGammaStack}; 
but for $h_2(r)$. Frame ($A$): $V_0=1.579/T_d$, $V_1=0.1/T_d$ and: $\gamma=16.667$ (thick-solid line); 
12.500 (dashed line); 6.250 (thin solid line); 3.125 (fine-dotted line); 2.500 (long-dashed dotted line); 
2.000 (short-dashed dotted line). Frame ($B$) is for $V_0=1.579/T_d$, $V_1=1.000/T_d$ and: $\gamma=10.000$
(solid line); 6.250 (dashed line); 4.000 (fine-dotted line); 3.333 (dashed-dotted line); 2.778 (double-dotted
line); 2.500 (thin solid line). Frame ($C$): $V_0=3.158$, $V_1=0.5/T_d$, and: $\gamma=25.000$ (thick-dashed line);
10.000 (thin-dashed line); 6.250 (dashed-dotted line); 1.724 (thick-solid line); 1.042 (circles).
}\label{fig:plotnormcorrGCANseveralAandBT0.001den0.400L500N200severalGammaStack}
\end{figure}

\section{Pair correlation function for inhomogeneous 1D Bose gases}\label{app:norm-corr-function-density}

\hs Fig.\ft\ref{fig:plotnormdensmnormcorrA1.579severalBasc2.000T0.001L500N200} is as in 
Fig.\ft\ref{fig:plotdensmcorrA1.579severalBasc2.000T0.001L500N200}; but for $h_2(r)$ given by 
Eq.(\ref{eq:normcorrh2r}). In frame $(A)$ for low $V_0$, the amplitude of the spatial oscillations in $h_2(r)$
is substantially reduced compared to the corresponding $g_2(r)$. Although the $h_2(r)-$oscillations do not 
vanish completely, their amplitude does not change with $V_1$ that is the same for $g_2(r)$ in 
Fig.\ft\ref{fig:plotdensmcorrA1.579severalBasc2.000T0.001L500N200}($A$). Indeed, the $g_2(r)-$oscillations 
are caused by the oscillations in the convoluted density $\rho_c(r)$ [Eq.(\ref{eq:spatially-varying-density-norm})], 
and a division of $g_2(r)n_0^2$ by $\rho_c(r)$ [to get $h_2(r)$] almost flattens $h_2(r)$ at larger $r$ so that 
it oscillates close to 1. Any changes in the structure of $h_2(r)$ with $V_1$ are not detected. There is a 
weak role of the band structure of the BCOL and its disorder at low $V_0$ which is consistent with 
Fig.\ft\ref{fig:plotdeltaandJ}. 

\hs As one increases $V_0$ in Fig.\ft\ref{fig:plotnormdensmnormcorrA1.579severalBasc2.000T0.001L500N200}($B$), 
the amplitude of oscillations in $h_2(r)$ rises similarly to $g_2(r)$. This time the normalization by $\rho_c(r)$
does not yield a significant suppression of the oscillations as it occurs in frame ($A$). Therefore, at larger 
$V_0$ the origin of these oscillations may not be anymore chiefly the oscillations in the spatial density; rather 
they largely arise from an increased effect of the band structure of the BCOL and the disorder; the latter is 
again consistent with the increase of disorder with $V_0$ in Fig.\ft\ref{fig:plotdeltaandJ}. Qualitatively 
nevertheless, Fig.\ft\ref{fig:plotnormdensmnormcorrA1.579severalBasc2.000T0.001L500N200} shows the same effects as 
Fig.\ft\ref{fig:plotdensmcorrA1.579severalBasc2.000T0.001L500N200}, that is to say, that by either $g_2(r)$ or 
$h_2(r)$, one reaches qualitatively the same conclusions about the effect of $V_1$ on the strength of the 
correlations, namely that there is none at low $V_0$.

\hs Fig.\ft\ref{fig:plotnormcorrGCANseveralAandBT0.001den0.400L500N200severalGammaStack} is the same as
Fig.\ft\ref{fig:plotcorrGCANseveralAandBT0.001den0.400L500N200severalGammaStack}; but for $h_2(r)$. Here,
the effect of interactions $\gamma$ at $r\stackrel{>}{~}0.00025$ is substantially weakened in by the division 
of $g_2(r)n_0^2$ by $\rho_c(r)$, but remains quite pronounced towards $r\rightarrow 0$. The effect of $\gamma$ at 
larger $r$ cannot therefore be significantly detected by $h_2(r)$ because the effect of interactions is absorbed 
by the normalization process. At $r=0$, $h_2(r)$ displays a reduction with $\gamma$ similar to 
Fig.\ft\ref{fig:plotcorrGCANseveralAandBT0.001den0.400L500N200severalGammaStack} and shows the same
decline as in Fig.\ft\ref{fig:compareAnalyticalcorr0withWAPIMC} (see Sec.\ref{app:local-h2(0)} below).

\section{One-body density matrix for inhomogeneous 1D Bose gases}\label{app:sa-obdm}

\hs The SA-OBDM $h_1(r)$ [Eq.(\ref{eq:obdm-inhomogeneous})] in frames ($C$) and ($D$) of 
Fig.\ft\ref{fig:plotnormdensmnormcorrA1.579severalBasc2.000T0.001L500N200}
displays qualitatively the same behavior as the corresponding $g_1(r)$, except for small oscillations
that are a result of normalizing by $\rho_{c,\frac{1}{2}}(r)$. In fact, at larger $V_0$, there is really
not much difference in the qualitative structure of $h_1(r)$ and $g_1(r)$ in frames ($D$) of 
Figs.\ft\ref{fig:plotnormdensmnormcorrA1.579severalBasc2.000T0.001L500N200} and 
\ref{fig:plotdensmcorrA1.579severalBasc2.000T0.001L500N200}, respectively $-$both of them display small oscillations.
Therefore for larger $V_0$, $h_1(r)$ and $g_1(r)$ can both be applied on the same
footage to draw conclusions about the superfluid depletion and role of the BCOL.

\section{Local correlation function for inhomogeneous 1D Bose gases}\label{app:local-h2(0)}

\begin{figure}
\includegraphics[width=8.5cm,bb=67 373 535 681,clip]{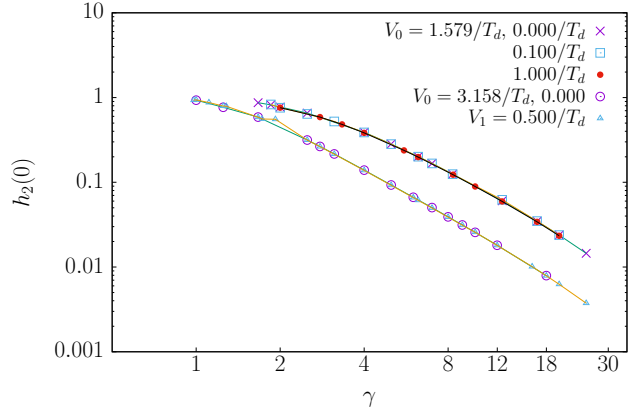}
\caption{As in Fig.\ft\ref{fig:compareAnalyticalcorr0withWAPIMC}; but for $h_2(0)$ with $V_0=1.579/T_d$
and: $V_1=0$ (times); $0.1/T_d$ (open square); $1.0/T_d$ (solid circles). Next data is for $V_0=3.158$ and:
$V_1=0$ (open circles) and $0.5/T_d$ (open triangles).}\label{fig:compareAnalyticalnormcorr0withWAPIMC}
\end{figure}

\hs The SAPCF at $r=0$, i.e. $h_2(0)$, displays in Fig.\ft\ref{fig:compareAnalyticalnormcorr0withWAPIMC} a 
drop with increasing $\gamma$ and shows almost the same behavior as $g_2(0)$ in 
Fig.\ft\ref{fig:compareAnalyticalcorr0withWAPIMC}. Again, changes in $V_1$ for the same $V_0$
have no effect on $h_2(0)$.

%\pagebreak
\bibliography{disorderedbosons,sfa,GBEC,optical-lattice}

\begin{thebibliography}{67}
\expandafter\ifx\csname natexlab\endcsname\relax\def\natexlab#1{#1}\fi
\expandafter\ifx\csname bibnamefont\endcsname\relax
  \def\bibnamefont#1{#1}\fi
\expandafter\ifx\csname bibfnamefont\endcsname\relax
  \def\bibfnamefont#1{#1}\fi
\expandafter\ifx\csname citenamefont\endcsname\relax
  \def\citenamefont#1{#1}\fi
\expandafter\ifx\csname url\endcsname\relax
  \def\url#1{\texttt{#1}}\fi
\expandafter\ifx\csname urlprefix\endcsname\relax\def\urlprefix{URL }\fi
\providecommand{\bibinfo}[2]{#2}
\providecommand{\eprint}[2][]{\url{#2}}

\bibitem[{\citenamefont{{P. Sengupta, M. Rigol, G. G. Batrouni, P. J. H.
  Denteneer, and R. T. Scalettar}}(2005)}]{Sengupta:2005}
\bibinfo{author}{\bibnamefont{{P. Sengupta, M. Rigol, G. G. Batrouni, P. J. H.
  Denteneer, and R. T. Scalettar}}}, \bibinfo{journal}{Phys. Rev. Lett.}
  \textbf{\bibinfo{volume}{95}}, \bibinfo{pages}{220402}
  (\bibinfo{year}{2005}).

\bibitem[{\citenamefont{{Xia-Ji Liu, Peter D. Drummond, and Hui
  Hu}}(2005)}]{Liu:2005}
\bibinfo{author}{\bibnamefont{{Xia-Ji Liu, Peter D. Drummond, and Hui Hu}}},
  \bibinfo{journal}{Phys. Rev. Lett.} \textbf{\bibinfo{volume}{94}},
  \bibinfo{pages}{136406} (\bibinfo{year}{2005}).

\bibitem[{\citenamefont{{S. R. Clark and D. Jaksch}}(2004)}]{Clark:2004}
\bibinfo{author}{\bibnamefont{{S. R. Clark and D. Jaksch}}},
  \bibinfo{journal}{Phys. Rev. A} \textbf{\bibinfo{volume}{70}},
  \bibinfo{pages}{043612} (\bibinfo{year}{2004}).

\bibitem[{\citenamefont{{S. Ramanan, T. Mishra, M. S. Luthra, R. V. Pai, and B.
  P. Das}}(2009)}]{Ramanan:2009}
\bibinfo{author}{\bibnamefont{{S. Ramanan, T. Mishra, M. S. Luthra, R. V. Pai,
  and B. P. Das}}}, \bibinfo{journal}{Phys. Rev. A}
  \textbf{\bibinfo{volume}{79}}, \bibinfo{pages}{013625}
  (\bibinfo{year}{2009}).

\bibitem[{\citenamefont{{Elmar Haller, Russel Hart, Manfred J. Mark, Johann G.
  Danzl, Lukas Reichs\"ollner, Mattias Gustavsson, Marcello Dalmonte, Guido
  Pupillo, and Hanns-Christoph N\"agerl}}(2010)}]{Haller:2010}
\bibinfo{author}{\bibnamefont{{Elmar Haller, Russel Hart, Manfred J. Mark,
  Johann G. Danzl, Lukas Reichs\"ollner, Mattias Gustavsson, Marcello Dalmonte,
  Guido Pupillo, and Hanns-Christoph N\"agerl}}}, \bibinfo{journal}{Nature}
  \textbf{\bibinfo{volume}{466}}, \bibinfo{pages}{597} (\bibinfo{year}{2010}).

\bibitem[{\citenamefont{{H. P. B\"uchler, G. Blatter, and W.
  Zwerger}}(2003)}]{Buechler:2003}
\bibinfo{author}{\bibnamefont{{H. P. B\"uchler, G. Blatter, and W. Zwerger}}},
  \bibinfo{journal}{Phys. Rev. Lett.} \textbf{\bibinfo{volume}{90}},
  \bibinfo{pages}{130401} (\bibinfo{year}{2003}).

\bibitem[{\citenamefont{{T. Giamarchi}}(2003)}]{Giamarchi:2003}
\bibinfo{author}{\bibnamefont{{T. Giamarchi}}}, \emph{\bibinfo{title}{{Quantum
  {P}hysics in {O}ne {D}imension}}} (\bibinfo{publisher}{{Oxford Univ. Press}},
  \bibinfo{year}{2003}), \bibinfo{edition}{1st} ed.

\bibitem[{\citenamefont{{Ming-Xia Huo and Dimitris G.
  Angelakis}}(2012)}]{Huo:2012}
\bibinfo{author}{\bibnamefont{{Ming-Xia Huo and Dimitris G. Angelakis}}},
  \bibinfo{journal}{Phys. Rev. A} \textbf{\bibinfo{volume}{85}},
  \bibinfo{pages}{023821} (\bibinfo{year}{2012}).

\bibitem[{\citenamefont{{E. H. Lieb and W. Liniger}}(1963)}]{Lieb:1963}
\bibinfo{author}{\bibnamefont{{E. H. Lieb and W. Liniger}}},
  \bibinfo{journal}{Phys. Rev.} \textbf{\bibinfo{volume}{130}},
  \bibinfo{pages}{1605} (\bibinfo{year}{1963}).

\bibitem[{\citenamefont{Guidoni et~al.}(1997)\citenamefont{Guidoni, Trich\'e,
  Verkerk, and Grynberg}}]{Guidoni:1997}
\bibinfo{author}{\bibfnamefont{L.}~\bibnamefont{Guidoni}},
  \bibinfo{author}{\bibfnamefont{C.}~\bibnamefont{Trich\'e}},
  \bibinfo{author}{\bibfnamefont{P.}~\bibnamefont{Verkerk}}, \bibnamefont{and}
  \bibinfo{author}{\bibfnamefont{G.}~\bibnamefont{Grynberg}},
  \bibinfo{journal}{Phys. Rev. Lett.} \textbf{\bibinfo{volume}{79}},
  \bibinfo{pages}{3363} (\bibinfo{year}{1997}).

\bibitem[{\citenamefont{{Nicolas Nessi and Anibal Iucci}}(2011)}]{Nessi:2011}
\bibinfo{author}{\bibnamefont{{Nicolas Nessi and Anibal Iucci}}},
  \bibinfo{journal}{Phys. Rev. A} \textbf{\bibinfo{volume}{84}},
  \bibinfo{pages}{063614} (\bibinfo{year}{2011}).

\bibitem[{\citenamefont{{M. C. Gordillo, C. Carbonell-Coronado, and F. De
  Soto}}(2015)}]{Gordillo:2015}
\bibinfo{author}{\bibnamefont{{M. C. Gordillo, C. Carbonell-Coronado, and F. De
  Soto}}}, \bibinfo{journal}{Phys. Rev. A} \textbf{\bibinfo{volume}{91}},
  \bibinfo{pages}{043618} (\bibinfo{year}{2015}).

\bibitem[{\citenamefont{{Mohammad Hafezi, Darrick E. Chang, Vladimir Gritsev,
  Eugene Demler, and Mikhail D. Lukin}}(2012)}]{Hafezi:2012}
\bibinfo{author}{\bibnamefont{{Mohammad Hafezi, Darrick E. Chang, Vladimir
  Gritsev, Eugene Demler, and Mikhail D. Lukin}}}, \bibinfo{journal}{Phys. Rev.
  A} \textbf{\bibinfo{volume}{85}}, \bibinfo{pages}{013822}
  (\bibinfo{year}{2012}).

\bibitem[{\citenamefont{Pethick and Smith}(2002)}]{Pethick:2002}
\bibinfo{author}{\bibfnamefont{C.~J.} \bibnamefont{Pethick}} \bibnamefont{and}
  \bibinfo{author}{\bibfnamefont{H.}~\bibnamefont{Smith}},
  \emph{\bibinfo{title}{{Bose-Einstein Condensation in Dilute Gases}}}
  (\bibinfo{publisher}{{Cambridge University Press}},
  \bibinfo{address}{{Cambridge UK}}, \bibinfo{year}{2002}),
  \bibinfo{edition}{1st} ed.

\bibitem[{\citenamefont{{Maciej Lewenstein, Anna Sanpera, and
  Ver$\acute{o}$nica Ahufinger}}(2012)}]{Lewenstein:2012}
\bibinfo{author}{\bibnamefont{{Maciej Lewenstein, Anna Sanpera, and
  Ver$\acute{o}$nica Ahufinger}}}, \emph{\bibinfo{title}{{Ultracold {A}toms in
  {O}ptical {L}attices: {\it {S}imulating {Q}uantum {M}any-{B}ody {S}ystems}}}}
  (\bibinfo{publisher}{{Oxford University Press}}, \bibinfo{address}{{United
  Kingdom}}, \bibinfo{year}{2012}).

\bibitem[{\citenamefont{{G. Bo$\acute{e}$ris, L. Gori, M. D. Hoogerland, A.
  Kumar, E. Lucioni, L. Tanzi, M. Inguscio, T. Giamarchi, C. D’Errico, G.
  Carleo, G. Modugno, and L. Sanchez-Palencia}}(2016)}]{Boeris:2016}
\bibinfo{author}{\bibnamefont{{G. Bo$\acute{e}$ris, L. Gori, M. D. Hoogerland,
  A. Kumar, E. Lucioni, L. Tanzi, M. Inguscio, T. Giamarchi, C. D’Errico, G.
  Carleo, G. Modugno, and L. Sanchez-Palencia}}}, \bibinfo{journal}{Phys. Rev.
  A} \textbf{\bibinfo{volume}{93}}, \bibinfo{pages}{011601(R)}
  (\bibinfo{year}{2016}).

\bibitem[{\citenamefont{{Toshiya Kinoshita, Trevor Wenger, and David S.
  Weiss}}(2005)}]{Kinoshita:2005}
\bibinfo{author}{\bibnamefont{{Toshiya Kinoshita, Trevor Wenger, and David S.
  Weiss}}}, \bibinfo{journal}{Phys. Rev. Lett.} \textbf{\bibinfo{volume}{95}},
  \bibinfo{pages}{190406} (\bibinfo{year}{2005}).

\bibitem[{\citenamefont{{B. Deissler, E. Lucioni, M. Modugno, G. Roati, L.
  Tanzi, M. Zaccanti, M. Inguscio, and G. Modugno}}(2011)}]{Deissler:2011}
\bibinfo{author}{\bibnamefont{{B. Deissler, E. Lucioni, M. Modugno, G. Roati,
  L. Tanzi, M. Zaccanti, M. Inguscio, and G. Modugno}}}, \bibinfo{journal}{New
  J. Phys.} \textbf{\bibinfo{volume}{13}}, \bibinfo{pages}{023020}
  (\bibinfo{year}{2011}).

\bibitem[{\citenamefont{{G. Roux, T. Barthel, I. P. McCulloch, C. Kollath, U.
  Schollw\"ock, and T. Giamarchi}}(2008)}]{Roux:2008}
\bibinfo{author}{\bibnamefont{{G. Roux, T. Barthel, I. P. McCulloch, C.
  Kollath, U. Schollw\"ock, and T. Giamarchi}}}, \bibinfo{journal}{Phys. Rev.
  A} \textbf{\bibinfo{volume}{78}}, \bibinfo{pages}{023628}
  (\bibinfo{year}{2008}).

\bibitem[{\citenamefont{{Tommaso Roscilde}}(2008)}]{Roscilde:2008}
\bibinfo{author}{\bibnamefont{{Tommaso Roscilde}}}, \bibinfo{journal}{Phys.
  Rev. A} \textbf{\bibinfo{volume}{77}}, \bibinfo{pages}{063605}
  (\bibinfo{year}{2008}).

\bibitem[{\citenamefont{{M. Larcher, M. Modugno, and F.
  Dalfovo}}(2011)}]{Larcher:2011}
\bibinfo{author}{\bibnamefont{{M. Larcher, M. Modugno, and F. Dalfovo}}},
  \bibinfo{journal}{Phys. Rev. A} \textbf{\bibinfo{volume}{83}},
  \bibinfo{pages}{013624} (\bibinfo{year}{2011}).

\bibitem[{\citenamefont{{Michele Modugno}}(2009)}]{Modugno:2009}
\bibinfo{author}{\bibnamefont{{Michele Modugno}}}, \bibinfo{journal}{New J.
  Phys.} \textbf{\bibinfo{volume}{11}}, \bibinfo{pages}{033023}
  (\bibinfo{year}{2009}).

\bibitem[{\citenamefont{{R. Roth and K. Burnett}}(2003)}]{Roth:2003}
\bibinfo{author}{\bibnamefont{{R. Roth and K. Burnett}}},
  \bibinfo{journal}{Phys. Rev. A} \textbf{\bibinfo{volume}{67}},
  \bibinfo{pages}{031602(R)} (\bibinfo{year}{2003}).

\bibitem[{\citenamefont{{S. Pilati, S. Giorgini, M. Modugno, and N.
  Prokof'ev}}(2010)}]{Pilati:2010}
\bibinfo{author}{\bibnamefont{{S. Pilati, S. Giorgini, M. Modugno, and N.
  Prokof'ev}}}, \bibinfo{journal}{New J. Phys.} \textbf{\bibinfo{volume}{12}},
  \bibinfo{pages}{073003} (\bibinfo{year}{2010}).

\bibitem[{\citenamefont{{I. L. Aleiner, B. L. Altshuler, and G. V.
  Shlyapnikov}}(2010)}]{Aleiner:2010}
\bibinfo{author}{\bibnamefont{{I. L. Aleiner, B. L. Altshuler, and G. V.
  Shlyapnikov}}}, \bibinfo{journal}{Nature Phys.} \textbf{\bibinfo{volume}{6}},
  \bibinfo{pages}{900} (\bibinfo{year}{2010}).

\bibitem[{\citenamefont{{P. Lugan, A. Aspect, L. Sanchez-Palencia, D. Delande,
  B. Gr$\acute{e}$maud, C. A. M\"uller, and C. Miniatura}}(2009)}]{Lugan:2009}
\bibinfo{author}{\bibnamefont{{P. Lugan, A. Aspect, L. Sanchez-Palencia, D.
  Delande, B. Gr$\acute{e}$maud, C. A. M\"uller, and C. Miniatura}}},
  \bibinfo{journal}{Phys. Rev. A} \textbf{\bibinfo{volume}{80}},
  \bibinfo{pages}{023605} (\bibinfo{year}{2009}).

\bibitem[{\citenamefont{{M. White, M. Pasienski, D. McKay, S. Q. Zhou, D.
  Ceperley, and B. DeMarco}}(2009)}]{White:2009}
\bibinfo{author}{\bibnamefont{{M. White, M. Pasienski, D. McKay, S. Q. Zhou, D.
  Ceperley, and B. DeMarco}}}, \bibinfo{journal}{Phys. Rev. Lett.}
  \textbf{\bibinfo{volume}{102}}, \bibinfo{pages}{055301}
  (\bibinfo{year}{2009}).

\bibitem[{\citenamefont{{L. Fallani, J. E. Lye, V. Guarrera, C. Fort, and M.
  Inguscio}}(2007)}]{Fallani:2007}
\bibinfo{author}{\bibnamefont{{L. Fallani, J. E. Lye, V. Guarrera, C. Fort, and
  M. Inguscio}}}, \bibinfo{journal}{Phys. Rev. Lett.}
  \textbf{\bibinfo{volume}{98}}, \bibinfo{pages}{130404}
  (\bibinfo{year}{2007}).

\bibitem[{\citenamefont{{B. Deissler, M. Zaccanti, G. Roati, C. D\'Errico, M.
  Fattori, M. Modugno, G. Modugno, and M. Inguscio}}(2010)}]{Deissler:2010}
\bibinfo{author}{\bibnamefont{{B. Deissler, M. Zaccanti, G. Roati, C.
  D\'Errico, M. Fattori, M. Modugno, G. Modugno, and M. Inguscio}}},
  \bibinfo{journal}{Nature Phys.} \textbf{\bibinfo{volume}{6}},
  \bibinfo{pages}{354} (\bibinfo{year}{2010}).

\bibitem[{\citenamefont{{Giacomo Roati, Chiara D\'Errico, Leonardo Fallani,
  Marco Fattori, Chiara Fort, Matteo Zaccanti, Giovanni Modugno, Michele
  Modugno, and Massimo Inguscio}}(2008)}]{Roati:2008}
\bibinfo{author}{\bibnamefont{{Giacomo Roati, Chiara D\'Errico, Leonardo
  Fallani, Marco Fattori, Chiara Fort, Matteo Zaccanti, Giovanni Modugno,
  Michele Modugno, and Massimo Inguscio}}}, \bibinfo{journal}{Nature}
  \textbf{\bibinfo{volume}{453}}, \bibinfo{pages}{895} (\bibinfo{year}{2008}).

\bibitem[{\citenamefont{{Juliette Billy, Vincent Josse, Zhanchun Zuo, Alain
  Bernard, Ben Hambrecht, Pierre Lugan, David Cl$\acute{e}$ment, Laurent
  Sanchez-Palencia, Philippe Bouyer, and Alain Aspect}}(2008)}]{Billy:2008}
\bibinfo{author}{\bibnamefont{{Juliette Billy, Vincent Josse, Zhanchun Zuo,
  Alain Bernard, Ben Hambrecht, Pierre Lugan, David Cl$\acute{e}$ment, Laurent
  Sanchez-Palencia, Philippe Bouyer, and Alain Aspect}}},
  \bibinfo{journal}{Nature Phys. Lett.} \textbf{\bibinfo{volume}{453}},
  \bibinfo{pages}{891} (\bibinfo{year}{2008}).

\bibitem[{\citenamefont{{Yong P. Chen, J. Hitchcock, D. Dries, M. Junker, C.
  Welford, and R. G. Hulet}}(2008)}]{Chen:2008}
\bibinfo{author}{\bibnamefont{{Yong P. Chen, J. Hitchcock, D. Dries, M. Junker,
  C. Welford, and R. G. Hulet}}}, \bibinfo{journal}{Phys. Rev. A}
  \textbf{\bibinfo{volume}{77}}, \bibinfo{pages}{033632}
  (\bibinfo{year}{2008}).

\bibitem[{\citenamefont{{Matthew P. A. Fisher, Peter B. Weichman, G. Grinstein,
  and Daniel S. Fisher}}(1989)}]{Fisher:1989}
\bibinfo{author}{\bibnamefont{{Matthew P. A. Fisher, Peter B. Weichman, G.
  Grinstein, and Daniel S. Fisher}}}, \bibinfo{journal}{Phys. Rev. B}
  \textbf{\bibinfo{volume}{40}}, \bibinfo{pages}{546} (\bibinfo{year}{1989}).

\bibitem[{\citenamefont{{Jacques Bossy, Jonathan V. Pearce, Helmut Schober, and
  Henry R. Glyde}}(2008)}]{Bossy:2008}
\bibinfo{author}{\bibnamefont{{Jacques Bossy, Jonathan V. Pearce, Helmut
  Schober, and Henry R. Glyde}}}, \bibinfo{journal}{Phys. Rev. B}
  \textbf{\bibinfo{volume}{78}}, \bibinfo{pages}{224507}
  (\bibinfo{year}{2008}).

\bibitem[{\citenamefont{{S. Palpacelli and S. Succi}}(2008)}]{Palpacelli:2008}
\bibinfo{author}{\bibnamefont{{S. Palpacelli and S. Succi}}},
  \bibinfo{journal}{Phys. Rev. E} \textbf{\bibinfo{volume}{77}},
  \bibinfo{pages}{066708} (\bibinfo{year}{2008}).

\bibitem[{\citenamefont{{D. Cl$\acute{e}$ment, A. F. Var$\acute{o}$n, J. A.
  Retter, L. Sanchez-Palencia, A. Aspect, and P. Bouyer}}(2006)}]{Clement:2006}
\bibinfo{author}{\bibnamefont{{D. Cl$\acute{e}$ment, A. F. Var$\acute{o}$n, J.
  A. Retter, L. Sanchez-Palencia, A. Aspect, and P. Bouyer}}},
  \bibinfo{journal}{New J. Phys.} \textbf{\bibinfo{volume}{8}},
  \bibinfo{pages}{165} (\bibinfo{year}{2006}).

\bibitem[{\citenamefont{Deng et~al.}(2013)\citenamefont{Deng, Citro, Orignac,
  Minguzzi, and Santos}}]{Deng:2013}
\bibinfo{author}{\bibfnamefont{X.}~\bibnamefont{Deng}},
  \bibinfo{author}{\bibfnamefont{R.}~\bibnamefont{Citro}},
  \bibinfo{author}{\bibfnamefont{E.}~\bibnamefont{Orignac}},
  \bibinfo{author}{\bibfnamefont{A.}~\bibnamefont{Minguzzi}}, \bibnamefont{and}
  \bibinfo{author}{\bibfnamefont{L.}~\bibnamefont{Santos}},
  \bibinfo{journal}{Phys. Rev. B} \textbf{\bibinfo{volume}{87}},
  \bibinfo{pages}{195101} (\bibinfo{year}{2013}).

\bibitem[{\citenamefont{Pollet et~al.}(2013)\citenamefont{Pollet, Prokof'ev,
  and Svistunov}}]{Pollet:2013}
\bibinfo{author}{\bibfnamefont{L.}~\bibnamefont{Pollet}},
  \bibinfo{author}{\bibfnamefont{N.~V.} \bibnamefont{Prokof'ev}},
  \bibnamefont{and} \bibinfo{author}{\bibfnamefont{B.~V.}
  \bibnamefont{Svistunov}}, \bibinfo{journal}{Phys. Rev. B}
  \textbf{\bibinfo{volume}{87}}, \bibinfo{pages}{144203}
  (\bibinfo{year}{2013}).

\bibitem[{\citenamefont{Ristivojevic et~al.}(2012)\citenamefont{Ristivojevic,
  Petkovi\ifmmode~\acute{c}\else \'{c}\fi{}, Le~Doussal, and
  Giamarchi}}]{Ristivojevic:2012}
\bibinfo{author}{\bibfnamefont{Z.}~\bibnamefont{Ristivojevic}},
  \bibinfo{author}{\bibfnamefont{A.}~\bibnamefont{Petkovi\ifmmode~\acute{c}\else
  \'{c}\fi{}}}, \bibinfo{author}{\bibfnamefont{P.}~\bibnamefont{Le~Doussal}},
  \bibnamefont{and}
  \bibinfo{author}{\bibfnamefont{T.}~\bibnamefont{Giamarchi}},
  \bibinfo{journal}{Phys. Rev. Lett.} \textbf{\bibinfo{volume}{109}},
  \bibinfo{pages}{026402} (\bibinfo{year}{2012}).

\bibitem[{\citenamefont{{Shankar Iyer, David Pekker, and Gil
  Rafael}}(2013)}]{Iyer:2013}
\bibinfo{author}{\bibnamefont{{Shankar Iyer, David Pekker, and Gil Rafael}}},
  \bibinfo{journal}{Phys. Rev. B} \textbf{\bibinfo{volume}{88}},
  \bibinfo{pages}{220501(R)} (\bibinfo{year}{2013}).

\bibitem[{\citenamefont{Basko and Hekking}(2013)}]{Basko:2013}
\bibinfo{author}{\bibfnamefont{D.~M.} \bibnamefont{Basko}} \bibnamefont{and}
  \bibinfo{author}{\bibfnamefont{F.~W.~J.} \bibnamefont{Hekking}},
  \bibinfo{journal}{Phys. Rev. B} \textbf{\bibinfo{volume}{88}},
  \bibinfo{pages}{094507} (\bibinfo{year}{2013}).

\bibitem[{\citenamefont{Schulte et~al.}(2005)\citenamefont{Schulte,
  Drenkelforth, Kruse, Ertmer, Arlt, Sacha, Zakrzewski, and
  Lewenstein}}]{Schulte:2005}
\bibinfo{author}{\bibfnamefont{T.}~\bibnamefont{Schulte}},
  \bibinfo{author}{\bibfnamefont{S.}~\bibnamefont{Drenkelforth}},
  \bibinfo{author}{\bibfnamefont{J.}~\bibnamefont{Kruse}},
  \bibinfo{author}{\bibfnamefont{W.}~\bibnamefont{Ertmer}},
  \bibinfo{author}{\bibfnamefont{J.}~\bibnamefont{Arlt}},
  \bibinfo{author}{\bibfnamefont{K.}~\bibnamefont{Sacha}},
  \bibinfo{author}{\bibfnamefont{J.}~\bibnamefont{Zakrzewski}},
  \bibnamefont{and}
  \bibinfo{author}{\bibfnamefont{M.}~\bibnamefont{Lewenstein}},
  \bibinfo{journal}{Phys. Rev. Lett.} \textbf{\bibinfo{volume}{95}},
  \bibinfo{pages}{170411} (\bibinfo{year}{2005}).

\bibitem[{\citenamefont{Paul et~al.}(2007)\citenamefont{Paul, Schlagheck,
  Leboeuf, and Pavloff}}]{Paul:2007}
\bibinfo{author}{\bibfnamefont{T.}~\bibnamefont{Paul}},
  \bibinfo{author}{\bibfnamefont{P.}~\bibnamefont{Schlagheck}},
  \bibinfo{author}{\bibfnamefont{P.}~\bibnamefont{Leboeuf}}, \bibnamefont{and}
  \bibinfo{author}{\bibfnamefont{N.}~\bibnamefont{Pavloff}},
  \bibinfo{journal}{Phys. Rev. Lett.} \textbf{\bibinfo{volume}{98}},
  \bibinfo{pages}{210602} (\bibinfo{year}{2007}).

\bibitem[{\citenamefont{Sanchez-Palencia
  et~al.}(2007)\citenamefont{Sanchez-Palencia, Cl\'ement, Lugan, Bouyer,
  Shlyapnikov, and Aspect}}]{Sanchez-Palencia:2007}
\bibinfo{author}{\bibfnamefont{L.}~\bibnamefont{Sanchez-Palencia}},
  \bibinfo{author}{\bibfnamefont{D.}~\bibnamefont{Cl\'ement}},
  \bibinfo{author}{\bibfnamefont{P.}~\bibnamefont{Lugan}},
  \bibinfo{author}{\bibfnamefont{P.}~\bibnamefont{Bouyer}},
  \bibinfo{author}{\bibfnamefont{G.~V.} \bibnamefont{Shlyapnikov}},
  \bibnamefont{and} \bibinfo{author}{\bibfnamefont{A.}~\bibnamefont{Aspect}},
  \bibinfo{journal}{Phys. Rev. Lett.} \textbf{\bibinfo{volume}{98}},
  \bibinfo{pages}{210401} (\bibinfo{year}{2007}).

\bibitem[{\citenamefont{Lugan et~al.}(2007)\citenamefont{Lugan, Cl\'ement,
  Bouyer, Aspect, and Sanchez-Palencia}}]{Lugan:2007}
\bibinfo{author}{\bibfnamefont{P.}~\bibnamefont{Lugan}},
  \bibinfo{author}{\bibfnamefont{D.}~\bibnamefont{Cl\'ement}},
  \bibinfo{author}{\bibfnamefont{P.}~\bibnamefont{Bouyer}},
  \bibinfo{author}{\bibfnamefont{A.}~\bibnamefont{Aspect}}, \bibnamefont{and}
  \bibinfo{author}{\bibfnamefont{L.}~\bibnamefont{Sanchez-Palencia}},
  \bibinfo{journal}{Phys. Rev. Lett.} \textbf{\bibinfo{volume}{99}},
  \bibinfo{pages}{180402} (\bibinfo{year}{2007}).

\bibitem[{\citenamefont{Paul et~al.}(2009)\citenamefont{Paul, Albert,
  Schlagheck, Leboeuf, and Pavloff}}]{Paul:2009}
\bibinfo{author}{\bibfnamefont{T.}~\bibnamefont{Paul}},
  \bibinfo{author}{\bibfnamefont{M.}~\bibnamefont{Albert}},
  \bibinfo{author}{\bibfnamefont{P.}~\bibnamefont{Schlagheck}},
  \bibinfo{author}{\bibfnamefont{P.}~\bibnamefont{Leboeuf}}, \bibnamefont{and}
  \bibinfo{author}{\bibfnamefont{N.}~\bibnamefont{Pavloff}},
  \bibinfo{journal}{Phys. Rev. A} \textbf{\bibinfo{volume}{80}},
  \bibinfo{pages}{033615} (\bibinfo{year}{2009}).

\bibitem[{\citenamefont{Radi\ifmmode~\acute{c}\else \'{c}\fi{}
  et~al.}(2010)\citenamefont{Radi\ifmmode~\acute{c}\else \'{c}\fi{}, Ba\ifmmode
  \check{c}\else \v{c}\fi{}i\ifmmode~\acute{c}\else \'{c}\fi{},
  Juki\ifmmode~\acute{c}\else \'{c}\fi{}, Segev, and Buljan}}]{Radic:2010}
\bibinfo{author}{\bibfnamefont{J.}~\bibnamefont{Radi\ifmmode~\acute{c}\else
  \'{c}\fi{}}}, \bibinfo{author}{\bibfnamefont{V.}~\bibnamefont{Ba\ifmmode
  \check{c}\else \v{c}\fi{}i\ifmmode~\acute{c}\else \'{c}\fi{}}},
  \bibinfo{author}{\bibfnamefont{D.}~\bibnamefont{Juki\ifmmode~\acute{c}\else
  \'{c}\fi{}}}, \bibinfo{author}{\bibfnamefont{M.}~\bibnamefont{Segev}},
  \bibnamefont{and} \bibinfo{author}{\bibfnamefont{H.}~\bibnamefont{Buljan}},
  \bibinfo{journal}{Phys. Rev. A} \textbf{\bibinfo{volume}{81}},
  \bibinfo{pages}{063639} (\bibinfo{year}{2010}).

\bibitem[{\citenamefont{Cestari et~al.}(2010)\citenamefont{Cestari, Foerster,
  and Gusm\~ao}}]{Cestari:2010}
\bibinfo{author}{\bibfnamefont{J.~C.~C.} \bibnamefont{Cestari}},
  \bibinfo{author}{\bibfnamefont{A.}~\bibnamefont{Foerster}}, \bibnamefont{and}
  \bibinfo{author}{\bibfnamefont{M.~A.} \bibnamefont{Gusm\~ao}},
  \bibinfo{journal}{Phys. Rev. A} \textbf{\bibinfo{volume}{82}},
  \bibinfo{pages}{063634} (\bibinfo{year}{2010}).

\bibitem[{\citenamefont{{S. Iyer, D. Pekker, and G. Rafael}}(2012)}]{Iyer:2012}
\bibinfo{author}{\bibnamefont{{S. Iyer, D. Pekker, and G. Rafael}}},
  \bibinfo{journal}{Phys. Rev. B} \textbf{\bibinfo{volume}{85}},
  \bibinfo{pages}{094202} (\bibinfo{year}{2012}).

\bibitem[{\citenamefont{{C. Aulbach, A. Wobst, G. L. Ingold, P. H\"anggi and I.
  Varga}}(2004)}]{Aulbach:2004}
\bibinfo{author}{\bibnamefont{{C. Aulbach, A. Wobst, G. L. Ingold, P. H\"anggi
  and I. Varga}}}, \bibinfo{journal}{New J. Phys.}
  \textbf{\bibinfo{volume}{6}}, \bibinfo{pages}{70} (\bibinfo{year}{2004}).

\bibitem[{\citenamefont{{D. J. Boers, B. Goedeke, D. Hinrichs and M.
  Holthaus}}(2007)}]{Boers:2007}
\bibinfo{author}{\bibnamefont{{D. J. Boers, B. Goedeke, D. Hinrichs and M.
  Holthaus}}}, \bibinfo{journal}{Phys. Rev. A} \textbf{\bibinfo{volume}{75}},
  \bibinfo{pages}{063404} (\bibinfo{year}{2007}).

\bibitem[{\citenamefont{{Xiaolong Deng and Luis Santos}}(2014)}]{Deng:2014}
\bibinfo{author}{\bibnamefont{{Xiaolong Deng and Luis Santos}}},
  \bibinfo{journal}{Phys. Rev. A} \textbf{\bibinfo{volume}{89}},
  \bibinfo{pages}{033632} (\bibinfo{year}{2014}).

\bibitem[{\citenamefont{{M. Boninsegni, N. V. Prokof'ev, and B. V.
  Svistunov}}(2006)}]{Boninsegni:2006}
\bibinfo{author}{\bibnamefont{{M. Boninsegni, N. V. Prokof'ev, and B. V.
  Svistunov}}}, \bibinfo{journal}{Phys. Rev. E} \textbf{\bibinfo{volume}{74}},
  \bibinfo{pages}{036701} (\bibinfo{year}{2006}).

\bibitem[{\citenamefont{{A. G. Sykes, D. M. Gangardt, M. J. Davis, K. Viering,
  M. G. Raizen, and K.V. Kheruntsyan}}(2008)}]{Sykes:2008}
\bibinfo{author}{\bibnamefont{{A. G. Sykes, D. M. Gangardt, M. J. Davis, K.
  Viering, M. G. Raizen, and K.V. Kheruntsyan}}}, \bibinfo{journal}{Phys. Rev.
  Lett.} \textbf{\bibinfo{volume}{100}}, \bibinfo{pages}{160406}
  (\bibinfo{year}{2008}).

\bibitem[{Pro()}]{Prokofev:2011}
\bibinfo{note}{{Nikolay Prokofev, UMASS Amherst, Private Communication}}.

\bibitem[{\citenamefont{Feynman}(1998)}]{Feynman:1998}
\bibinfo{author}{\bibfnamefont{R.~P.} \bibnamefont{Feynman}},
  \emph{\bibinfo{title}{{Statistical Mechanics}}}
  (\bibinfo{publisher}{{Westview Press, Advanced Book Program}},
  \bibinfo{address}{{Boulder, Colorado}}, \bibinfo{year}{1998}).

\bibitem[{\citenamefont{{G. E. Astrakharchik, J. Boronat, J. Casulleras, and S.
  Giorgini}}(2005)}]{Astrakharchik:2005}
\bibinfo{author}{\bibnamefont{{G. E. Astrakharchik, J. Boronat, J. Casulleras,
  and S. Giorgini}}}, \bibinfo{journal}{Phys. Rev. Lett.}
  \textbf{\bibinfo{volume}{95}}, \bibinfo{pages}{190407}
  (\bibinfo{year}{2005}).

\bibitem[{Pet()}]{PethickEqn}
\bibinfo{note}{{see Eq.(13.37) on page 349 in Ref.\cite{Pethick:2002}}}.

\bibitem[{foo()}]{footnote1}
\bibinfo{note}{{The second-order correlation function for atoms in an OL is
  given by Lewenstein \ea\ \cite{Lewenstein:2012} on page 419 and is
  effectively the same as ours: \begin{displaymath}
  g^{(2)}(\vec{\mathbf{d}})\,=\,\frac{\int
  d\vec{\mathbf{r}}\langle\hat{n}(\vec{\mathbf{r}}+\vec{\mathbf{d}}/2)
  \hat{n}(\vec{\mathbf{r}}\,-\,\vec{\mathbf{d}}/2)\rangle} {\int
  d\mathbf{r}\langle\hat{n}(\vec{\mathbf{r}}+\vec{\mathbf{d}}/2)\rangle
  \langle\hat{n}(\vec{\mathbf{r}}-\vec{\mathbf{d}}/2)\rangle}\,-\,1,
  \label{eq:Lewenstein-g2d} \end{displaymath} where
  $\vec{\mathbf{d}}=\vec{\mathbf{r}}-\vec{\mathbf{r^\prime}}$ is the distance
  between two particles situated at $\vec{\mathbf{r}}$ and
  $\vec{\mathbf{r^\prime}}$, and $\hat{n}(r)$ is the density operator.}}

\bibitem[{\citenamefont{{M. Naraschewski and R. J.
  Glauber}}(1999)}]{Naraschewski:1999}
\bibinfo{author}{\bibnamefont{{M. Naraschewski and R. J. Glauber}}},
  \bibinfo{journal}{Phys. Rev. A} \textbf{\bibinfo{volume}{59}},
  \bibinfo{pages}{4595} (\bibinfo{year}{1999}).

\bibitem[{\citenamefont{{Adrian Del Maestro and Ian
  Affleck}}(2010)}]{DelMaestro:2010}
\bibinfo{author}{\bibnamefont{{Adrian Del Maestro and Ian Affleck}}},
  \bibinfo{journal}{Phys. Rev. B} \textbf{\bibinfo{volume}{82}},
  \bibinfo{pages}{060515(R)} (\bibinfo{year}{2010}).

\bibitem[{\citenamefont{{Grigory E. Astrakharchik, Konstantin V. Krutitsky,
  Maciej Lewenstein, and Ferran Mazzanti}}()}]{Astrakharchik:2015}
\bibinfo{author}{\bibnamefont{{Grigory E. Astrakharchik, Konstantin V.
  Krutitsky, Maciej Lewenstein, and Ferran Mazzanti}}},
  \emph{\bibinfo{title}{{One-dimensional {B}ose gas in optical lattices of
  arbitrary strength}}}, \bibinfo{note}{{cond-mat/1509.01424}}.

\bibitem[{\citenamefont{{Alexander Yu. Cherny and Joachim
  Brand}}(2009)}]{Cherny:2009}
\bibinfo{author}{\bibnamefont{{Alexander Yu. Cherny and Joachim Brand}}},
  \bibinfo{journal}{Phys. Rev. A} \textbf{\bibinfo{volume}{79}},
  \bibinfo{pages}{043607} (\bibinfo{year}{2009}).

\bibitem[{\citenamefont{{K. V. Kheruntsyan, D. M. Gangardt, P. D. Drummond, and
  G. V. Shlyapnikov}}(2003)}]{Kheruntsyan:2003}
\bibinfo{author}{\bibnamefont{{K. V. Kheruntsyan, D. M. Gangardt, P. D.
  Drummond, and G. V. Shlyapnikov}}}, \bibinfo{journal}{Phys. Rev. Lett.}
  \textbf{\bibinfo{volume}{91}}, \bibinfo{pages}{040403}
  (\bibinfo{year}{2003}).

\bibitem[{\citenamefont{{Stefan S. Natu and Erich J.
  Mueller}}(2013)}]{Natu:2013}
\bibinfo{author}{\bibnamefont{{Stefan S. Natu and Erich J. Mueller}}},
  \bibinfo{journal}{Phys. Rev. A} \textbf{\bibinfo{volume}{87}},
  \bibinfo{pages}{063616} (\bibinfo{year}{2013}).

\bibitem[{\citenamefont{{E. E. Edwards, M. Beeler, Tao Hong, and S. L.
  Rolston}}(2008)}]{Edwards:2008}
\bibinfo{author}{\bibnamefont{{E. E. Edwards, M. Beeler, Tao Hong, and S. L.
  Rolston}}}, \bibinfo{journal}{Phys. Rev. Lett.}
  \textbf{\bibinfo{volume}{101}}, \bibinfo{pages}{260402}
  (\bibinfo{year}{2008}).

\bibitem[{\citenamefont{{G. D. Mahan}}(1990)}]{Mahan:1990}
\bibinfo{author}{\bibnamefont{{G. D. Mahan}}},
  \emph{\bibinfo{title}{{Many-Particle Physics}}} (\bibinfo{publisher}{{Plenum
  Press}}, \bibinfo{address}{{New York}}, \bibinfo{year}{1990}).

\end{thebibliography}
\end{document}